\documentclass[journal=ancac3,manuscript=article]{achemso}

\usepackage{chemformula} 
\usepackage[T1]{fontenc} 
\usepackage{graphicx}
\usepackage{color}
\usepackage{siunitx}
\usepackage{dcolumn}
\usepackage{latexsym}
\usepackage[normalem]{ulem}
\usepackage{hyperref,amssymb}
\usepackage{url}
\usepackage[numbers]{natbib}
\usepackage{adjustbox}
\usepackage{tabularx}
\usepackage{gensymb}
\usepackage{color,soul}
\usepackage{graphicx}
\usepackage{verbatim}
\usepackage{multirow}
\usepackage{amsmath}
\newcommand{\beq}{\begin{eqnarray}}
\newcommand{\eeq}{\end{eqnarray}}
\usepackage{mathrsfs}
\usepackage{float,soul}
\usepackage{mathtools}
\usepackage{slashed}
\usepackage{physics}	
\usepackage{graphicx}   
\usepackage{epstopdf}
\usepackage{subfigure}  
\usepackage{bbold}
\usepackage{wasysym}
\usepackage{feynmp}
\usepackage{xr}
\newcommand{\angstrom}{\textup{\AA}}

\author{Yuanxi Yu}
\affiliation{School of Physics and Astronomy, Shanghai Jiao Tong University, Shanghai 200240, China}
\alsoaffiliation{Institute of Natural Sciences, Shanghai Jiao Tong University, Shanghai 200240, China}
\author{Sha Jin}
\affiliation{School of Physics and Astronomy, Shanghai Jiao Tong University, Shanghai 200240, China}
\alsoaffiliation{Wilczek Quantum Center, Shanghai Jiao Tong University, Shanghai 200240, China}
\alsoaffiliation{Shanghai Research Center for Quantum Sciences, Shanghai 201315, China}
\author{Xue Fan}
\affiliation{Shanghai National Center for Applied Mathematics(SJTU Center), MOE-LSC, Shanghai Jiao Tong University, Shanghai 200240, China}
\alsoaffiliation{Materials Genome Institute, Shanghai University, Shanghai 200444, China}
\alsoaffiliation{School of Materials Science and Engineering, Georgia Institute of Technology, Atlanta, GA 30332, United States of America}
\author{Mona Sarter}
\affiliation{Rutherford Appleton Laboratory, ISIS Neutron and Muon Facility, Science and Technology Facilities Council, Didcot OX11 0QX, United Kingdom}
\author{Dehong Yu}
\affiliation{Australian Nuclear Science and Technology Organisation, Lucas Heights, NSW 2232, Australia}
\author{Liang Hong}
\email{hongl3liang@sjtu.edu.cn }
\affiliation{School of Physics and Astronomy, Shanghai Jiao Tong University, Shanghai 200240, China}
\alsoaffiliation{Institute of Natural Sciences, Shanghai Jiao Tong University, Shanghai 200240, China}
\alsoaffiliation{Shanghai National Center for Applied Mathematics(SJTU Center), MOE-LSC, Shanghai Jiao Tong University, Shanghai 200240, China}
\alsoaffiliation{Zhangjiang Institute for Advanced Study, Shanghai Jiao Tong University, Shanghai 201203, China}
\author{Matteo Baggioli}
\email{b.matteo@sjtu.edu.cn}
\affiliation{School of Physics and Astronomy, Shanghai Jiao Tong University, Shanghai 200240, China}
\alsoaffiliation{Wilczek Quantum Center, Shanghai Jiao Tong University, Shanghai 200240, China}
\alsoaffiliation{Shanghai Research Center for Quantum Sciences, Shanghai 201315, China}

\title[An \textsf{achemso} demo]
  {Emergence of Debye scaling in the density of states of liquids under nanoconfinement}


\begin{document}

\begin{tocentry}

\includegraphics{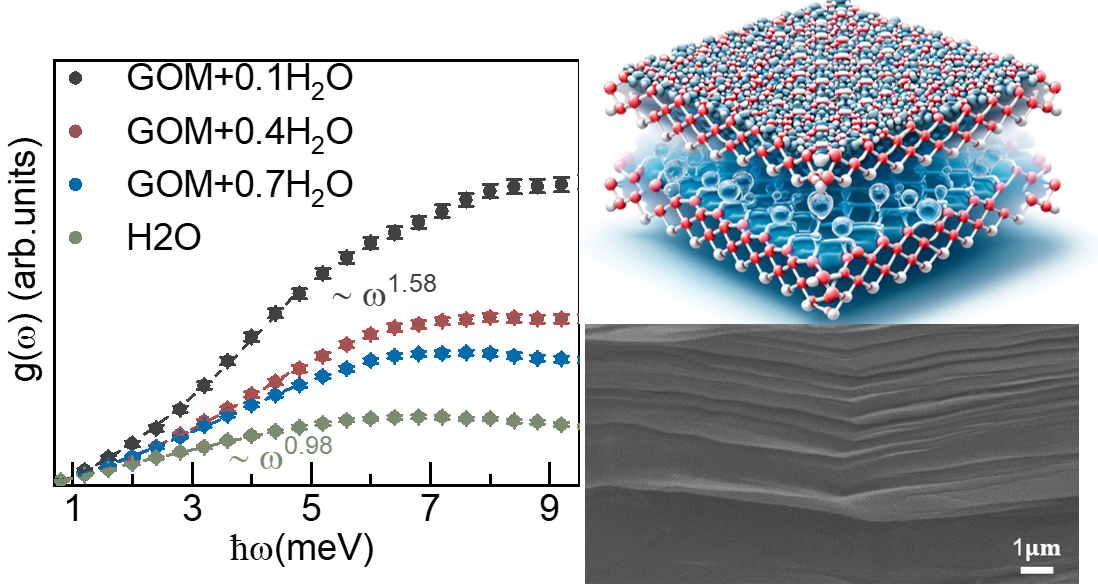}

\end{tocentry}

\begin{abstract}
In the realm of nanoscience, the dynamic behaviors of liquids at scales beyond the conventional structural relaxation time, $\tau$, unfold a fascinating blend of solid-like characteristics, including the propagation of collective shear waves and the emergence of elasticity. However, in classical bulk liquids, where $\tau$ is typically of the order of 1 ps or less, this solid-like behavior remains elusive in the low-frequency region of the density of states (\textit{DOS}). Here, we provide evidence for the emergent solid-like nature of liquids at short distances through inelastic neutron scattering measurements of the low-frequency DOS in liquid water and glycerol confined within graphene oxide membranes. In particular, upon increasing the strength of confinement, we observe a transition from a liquid-like \textit{DOS} (linear in the frequency $\omega$) to a solid-like behavior (Debye law, $\sim\omega^2$) in the range of $1$-$4$ meV. Molecular dynamics simulations confirm these findings and reveal additional solid-like features, including propagating collective shear waves and a reduction in the self-diffusion constant. Finally, we show that the onset of solid-like dynamics is pushed towards low frequency along with the slowing-down of the relaxation processes upon confinement. This nanoconfinement-induced transition, aligning with k-gap theory, underscores the potential of leveraging liquid nanoconfinement in advancing nanoscale science and technology, building more connections between fluid dynamics and materials engineering.
\end{abstract}


\newpage
From a structural point of view, liquids are profoundly different from solids. They do not display long-range order and they cannot be defined using the theoretical concept of spontaneous symmetry breaking and its direct consequences, such as the existence of long-wavelength phonon modes and the elasticity that emerges from that\cite{chaikin2000principles}. This poses fundamental questions in the description of the low-energy collective dynamics of liquids which are predominantly relaxational, rather than vibrational, and usually described with the theory of hydrodynamics\cite{boon1991molecular}.

It comes then as no surprise that the dynamical properties of liquids and solids at large distances and over long timescales -- the hydrodynamic regime -- are substantially different. To make this distinction more precise, it is instructive to introduce the concept of structural relaxation time $\tau$, which determines the speed of atomic re-arrangements in a liquid and the extent of the hydrodynamic regime. For bulk water at room temperature $\tau$ is of the order of $1$ ps \cite{PhysRevLett.82.775}, close to the value for the Maxwell relaxation time \cite{MALOMUZH2019111413,PhysRevFluids.4.123302}. In the rest of the manuscript, we will use the term low-frequency to indicate the range of energies below the inverse of the structural relaxation time for the bulk liquid, in which we do not expect any solid-like features. This corresponds to an energy of approximately $\hslash \omega < 4$meV for water at room temperature.

\begin{figure}[h]
     \centering
     \includegraphics[width=0.8\linewidth]{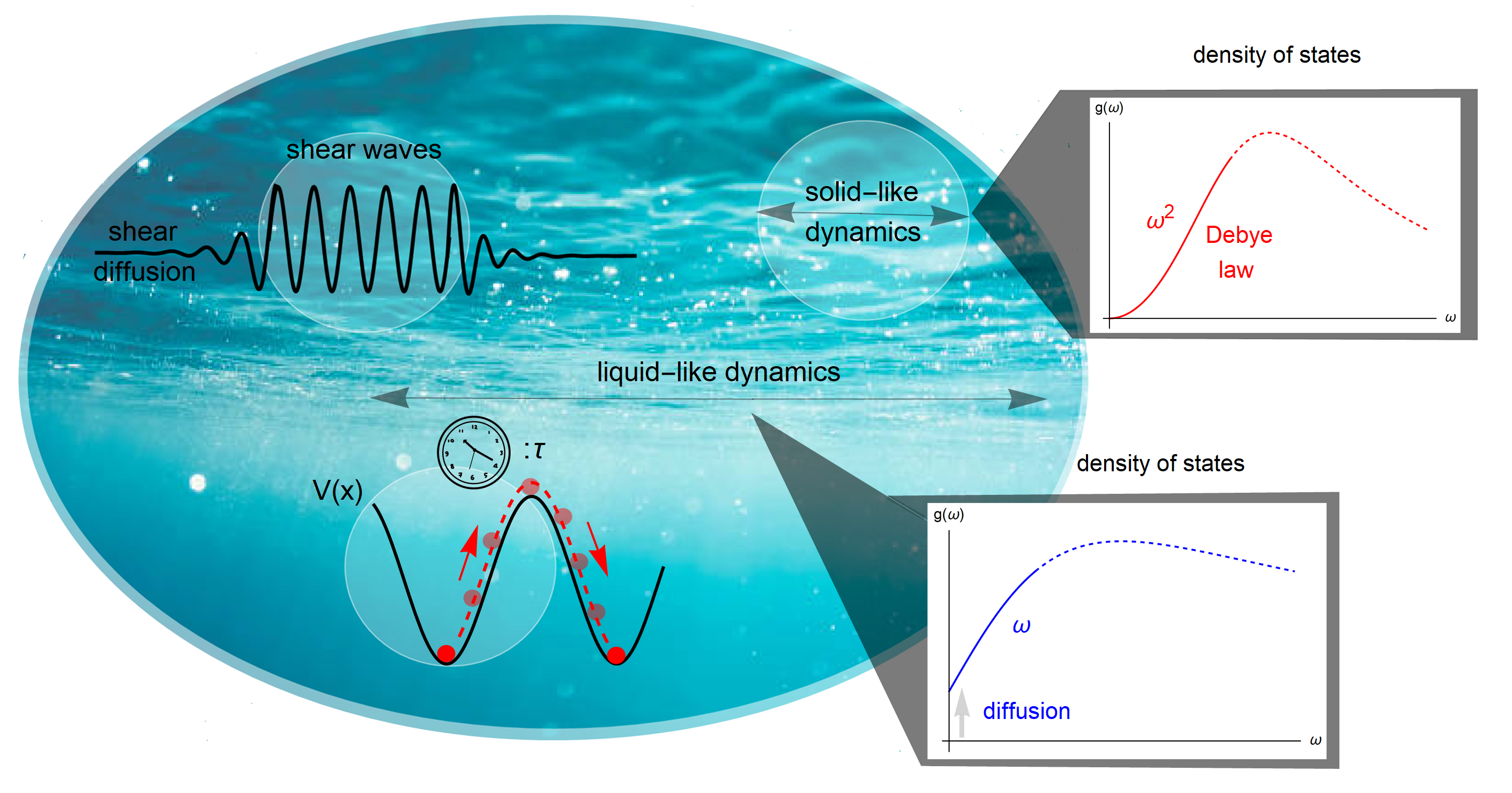}
     \caption{At short distances and short times, liquids exhibit solid-like properties. First, below a critical length-scale, propagating shear waves are expected in liquids, instead of the large wavelength shear diffusion. Second, for times below a relaxation timescale $\tau$, the dynamics are oscillatory and confined at the bottom of the potential landscape basin. Hence, the density of states of liquids is expected to be liquid-like at large scales and to become solid-like (Debye law) at short scales.}
     \label{fig:0}
 \end{figure}

A direct manifestation of the contrast between solids and liquids can be observed in the low-frequency behavior of their density of states (\textit{DOS}), $g(\omega)$. As a general statistical concept that measures the number of states per unit of energy within a system, \textit{DOS} can be applied to photons, electrons, phonons, and more. In the following discussion in this manuscript, the DOS we refer to is analogous to the concept of vibrational density of states (\textit{VDOS}) or phonon density of states found in solids and can be directly obtained from the dynamic structure factor. For liquids, the vibrational properties are less defined as for solids, especially at low energies, thus we simply use the term ``density of states'' (\textit{DOS}). 3D bulk solids display a characteristic quadratic scaling at low frequency $g(\omega)\propto \omega^2$, which is known as the Debye law\cite{kittel2018introduction}. Debye law indicates that the low-frequency dynamics of solids are governed by collective vibrational and propagating modes known as phonons, which can be identified with the Goldstone modes for the spontaneously broken translational symmetry\cite{Leutwyler:1996er}. On the contrary, the low-frequency \textit{DOS} of liquids exhibits two distinctive characteristics. First, the zero frequency value of $g(\omega)$ does not vanish and directly relates to the liquid relaxational dynamics, \textit{i.e.}, a non-zero self-diffusion constant\cite{hansen1990theory}. Second, the low-frequency scaling of the \textit{DOS} for 3D bulk liquids is linear in frequency, $g(\omega)\propto \omega$, as confirmed by neutron scattering experiments \cite{PhysRevLett.63.2381,DAWIDOWSKI2000247,dehong2022,jin2024temperaturedependencedensitystates,10.1063/1.469221}. For simplicity, in the rest of the manuscript, we will refer to the quadratic Debye behavior as ``solid-like'' and to the linear scaling in frequency as ``liquid-like'' (see insets in Figure \ref{fig:0}). We emphasize that in the case of liquids the linear scaling does not extrapolate all the way down to zero frequency, since the density of states remains finite at $\omega=0$, but it nevertheless appears in the ``low-frequency'' regime, corresponding to energies at which solids display a universal Debye scaling. We also avoid all complications regarding amorphous systems; hence, in the whole manuscript we always refer to crystalline solids.

After acknowledging that the low-frequency dynamics of bulk liquids are drastically different from those in solids, a  question arises as whether this sharp contrast persists at high frequencies or short scales.
It has been early realized by Frenkel\cite{frenkel} (see also Zwanzig \cite{PhysRev.156.190}, or the more recent reviews) \cite{trachenko2015collective,BAGGIOLI20201} that liquid dynamics can be qualitatively modeled as solid-like oscillatory motion around a position of local equilibrium interrupted by diffusive jumps toward different potential minima. These jumps constitute the fundamental origin of fluidity (liquid flow) and they happen at an average rate $1/\tau$ (see Figure \ref{fig:0}). As a consequence, the dynamics of liquids are expected to be solid-like for frequencies faster than this relaxation rate, $\omega \gg 1/\tau$, simply because relaxation has no time to take place. In other words, one does expect the appearance of solid-like vibrational modes in the high-frequency dynamics of liquids. This is indeed the case, as reported by many works \cite{PhysRevLett.32.49,hosokawa2009transverse,doi:10.1073/pnas.1006319107,Hosokawa_2015,simeoni2010widom,RevModPhys.77.881,doi:10.1080/14786435.2015.1096975}.

Using a continuum description, another way to model a transition between a propagating oscillatory frequency behavior and liquid-like diffusive dynamics is by using the so-called telegrapher equation, or k-gap theory \cite{BAGGIOLI20201}:
\begin{equation}\label{kgapeq}
    \omega^2+i \omega \tau_{g}^{-1}=v^2 k^2,
\end{equation}
where $v$ is the asymptotic speed of propagation of the collective waves at $\omega \gg 1/\tau_g$, and $k,\omega$ are respectively the wave-vector and the frequency of the collective mode.
In the expression above, that is valid for transverse excitations, $\tau_{g}$ is a timescale governing relaxation processes and expected to decrease with increasing temperature. By assuming Maxwell interpolation, one can identify $\tau_g$ with the Maxwell relaxation time $\tau_M \equiv \eta/G_{\infty}$, where $\eta$ and $G_{\infty}$ are respectively the shear viscosity and the instantaneous shear modulus \cite{trachenko2015collective,BAGGIOLI20201}. In principle, the average particle jump time $\tau$ does not have to quantitatively equate the collective Maxwell time $\tau_M$. Nevertheless, it has been explicitly proved that the Maxwell time agrees very well with the average lifetime of local atomic connectivity, that represents the characteristic timescale for atoms to lose or gain one nearest neighbor \cite{PhysRevLett.110.205504}. Given the several subtleties connected with the definition of this timescale, at this stage we will consider $\tau_g$ as an independent phenomenological parameter which can be extracted from the dispersion relation of gapped shear waves in liquids (see for example Ref.\citenum{PhysRevLett.118.215502}), and we will not make any further assumption.

Back to Eq.\eqref{kgapeq}, in the low frequency limit, the first term can be neglected and the dynamics result purely diffusive (liquid-like), with macroscopic diffusion constant $\mathcal{D}=v^2 \tau_{g}$ (not to be confused with the single-particle self-diffusion constant $D$). In the opposite limit, the second term can be neglected and the dynamics are purely vibrational and solid-like. We notice that Eq.\eqref{kgapeq} stops to be valid at very large $k$, as it is derived in a perturbative expansion in spatial gradient ($\sim k$). Moreover, a more careful analysis reveals still a crucial difference between the large frequency behavior arising from Eq.\eqref{kgapeq} and the dynamics of phonons in solids. Indeed, expanding the solution of Eq.\eqref{kgapeq} at large frequency, we obtain a dispersion of the type $\omega=v k -i/2\tau_g$, whose imaginary part is wave-vector independent and therefore qualitatively different from that of phonons in solids \cite{chaikin2000principles} (where at finite temperature and low wave-vector is quadratic in $k$, \textit{i.e.}, Akhiezer damping). This distinction is compatible with the observed broadening of high-frequency solid-like modes in liquids, which arises physically from the distribution of the liquid local structures \cite{doi:10.1073/pnas.1006319107}.

Eq.\eqref{kgapeq} appears in the description of several physical systems\cite{BAGGIOLI20201} (\textit{e.g.}, Cattaneo heat conduction equation) and plays a fundamental role in the understanding of the dynamics of collective shear waves in liquids, where it has been relabeled as the k-gap equation\cite{trachenko2015collective}. Besides modelling the high-frequency solid-like collective motion in liquids, Eq.\eqref{kgapeq} presents another striking prediction. In particular, it suggests the existence of shear waves with finite real frequency in liquids above a certain wave-vector cutoff, known as k-gap, and given by $k_g=1/(2 v \tau_{g})$. This prediction has been verified in simulations of classical liquids\cite{PhysRevLett.118.215502,PhysRevB.101.214312,Fomin_2020,PhysRevE.107.014139,doi:10.1021/acs.jpclett.9b03568,PhysRevLett.125.125501} and plasmas\cite{10.1063/1.5088141,PhysRevE.85.066401,PhysRevLett.85.2514,PhysRevLett.84.6026}, but experimentally confirmed only in a two-dimensional Yukawa dusty plasma \cite{PhysRevLett.97.115001} and in granular fluids \cite{jiang2024experimental}. In the literature, it is often claimed that the presence of a k-gap is tantamount to saying that below a critical length-scale $L_g=2 \pi/k_g$, and independently of the value of the frequency, the collective dynamics of liquids are akin to that of solids as they present propagating shear waves. Nevertheless, a more conservative approach requires that the real part of the frequency is at least larger than its imaginary part, or in other words, that the wave is able to propagate over a few wave-lengths before decaying. This second criterion applied to the k-gap equation qualitatively agrees with Frenkel original argument $\omega >1/\tau_g$ (see Supporting Information), if one identifies $\tau$ with $\tau_g$. This liquid-like to overdamped to propagating dynamics are in qualitative agreement with the experimental analysis for water presented in Ref.\citenum{PhysRevLett.79.1678}. There, it has been shown that the dynamics of room temperature water are liquid-like (hydrodynamic) for $q<2$nm$^{-1}$ (to be compared with our value for $k_g$ given by $\approx 2.2$nm$^{-1}$), overdamped in the range $2$nm$^{-1}<q<4$nm$^{-1}$, and solid-like above $q=4$nm$^{-1}$.

In parallel to this discussion, in recent years there has been intense progress in revealing and understanding the emergent short-scale elasticity of liquids \cite{10.1063/5.0051587,Noirez2021,doi:10.1073/pnas.2010787117}, which has ultimately led to the question of where the hydrodynamic limit really is\cite{doi:10.1080/08927022.2021.1975038,PhysRevD.103.086001}. The solid-like vibrational dynamics at short-scale are obviously connected to the emergence of macroscopic rigidity at short distances. A large part of this research program has been focused on disclosing the solid-like nature of liquids by considering confined systems, and therefore directly probing their short-scale dynamics.  Confined liquid systems include liquids in nanoporous materials, droplets, liquid films, and liquids affected by interfacial interactions on the microscopic scale. Under these circumstances, the interactions between atoms or molecules in confined liquids undergo significant changes due to geometric constraints, resulting in distinct manifestations in the \textit{DOS}. The effects of confinement on liquid structure, transport and dynamics have been explored in several works, specially for the case of water\cite{Tripathy2019,CALERO2020114027,PhysRevLett.100.106102,PhysRevLett.105.106101,Corti2021,KUO2023140612}. 

In confined liquids, the mechanical relaxation times have been shown to increase dramatically as compared to bulk behavior\cite{PhysRevLett.105.106101}. A strong enhancement of the thermal conductivity has also been observed under confinement \cite{Frank2017} and attributed, following Frenkel's idea \cite{frenkel}, to the presence of additional transverse oscillations occurring because of the larger liquid relaxation time. More closely to the topic of this manuscript, the low-energy behavior of the \textit{DOS} of confined water has been investigated in Ref.\citenum{ijms20215373} and more recently in Ref.\citenum{KUO2023140612}. In Ref.\citenum{ijms20215373} a distinctive difference between fragile confined water, where $g(\omega)\propto \omega^{1.5}$, and strong water, where $g(\omega)\propto \omega^2$, was observed and explained using mode-coupling theory and the idea of a crossover between heteoregeneity dominated dynamics to phonon-like excitations. In Ref.\cite{KUO2023140612}, a reduction of the fraction of unstable instantaneous normal modes and of the diffusion constant as a function of the confinement width, indicating the solid-like emergent dynamics, was reported. More importantly, the low-frequency power-law of the instantaneous normal mode \textit{DOS} has been studied as a function of the width and the direction.  Interestingly, Kuo \textit{et al}.\cite{KUO2023140612} found that the power-law in the direction of confinement, which is perpendicular to the confinement surface, grows gradually by reducing the width and reaches a value around $\approx 1.7$, close to Debye law. These recent findings will nicely resonate with our experimental results.

In summary, despite the effects of confinement on the structure and dynamics of liquids being largely explored in the past, a complete experimental proof of the emergence of a solid-like collective dynamics under confinement is still missing. In this work, we employed a combination of inelastic neutron scattering and molecular dynamics simulations to analyze the low-frequency (below $4$ meV) \textit{DOS} in liquid water and glycerol nano-confined within graphene oxide membranes (\textit{GOM}). \textit{GOM}, prepared by the modified Hummer's method \cite{marcano2010improved}, have $28\%$ oxidation and hydrophilic interfaces, enabling water and glycerol molecules easy access to the interlayers. Utilizing such techniques, we provide experimental evidence of the evolution of the scaling for the low-frequency \textit{DOS} in these two systems from a linear to quadratic Debye-like behavior by decreasing the confinement size. The validation of this phenomenon in two different systems suggests that emergent solid-like vibrational modes induced by confinement may be a universal feature of confined liquids, consistent with the observation of elastic response previously reported in the literature \cite{gao2007structured,Riedo}. Various microscopic quantities of the liquids, including current correlation functions, transverse and longitudinal wave dispersions, and self-diffusion constants are analyzed in order to characterize this finding in more detail. Exploiting qualitative arguments based on the k-gap theory \cite{BAGGIOLI20201}, we attribute the emergence of solid-like \textit{DOS} to the slowing down of relaxation processes under nanoconfinement, which corroborates the original perspective of liquid dynamics by Frenkel and Maxwell.

\section{Results and discussion}

\begin{figure*}[!t]\centering
\includegraphics[width=\linewidth]{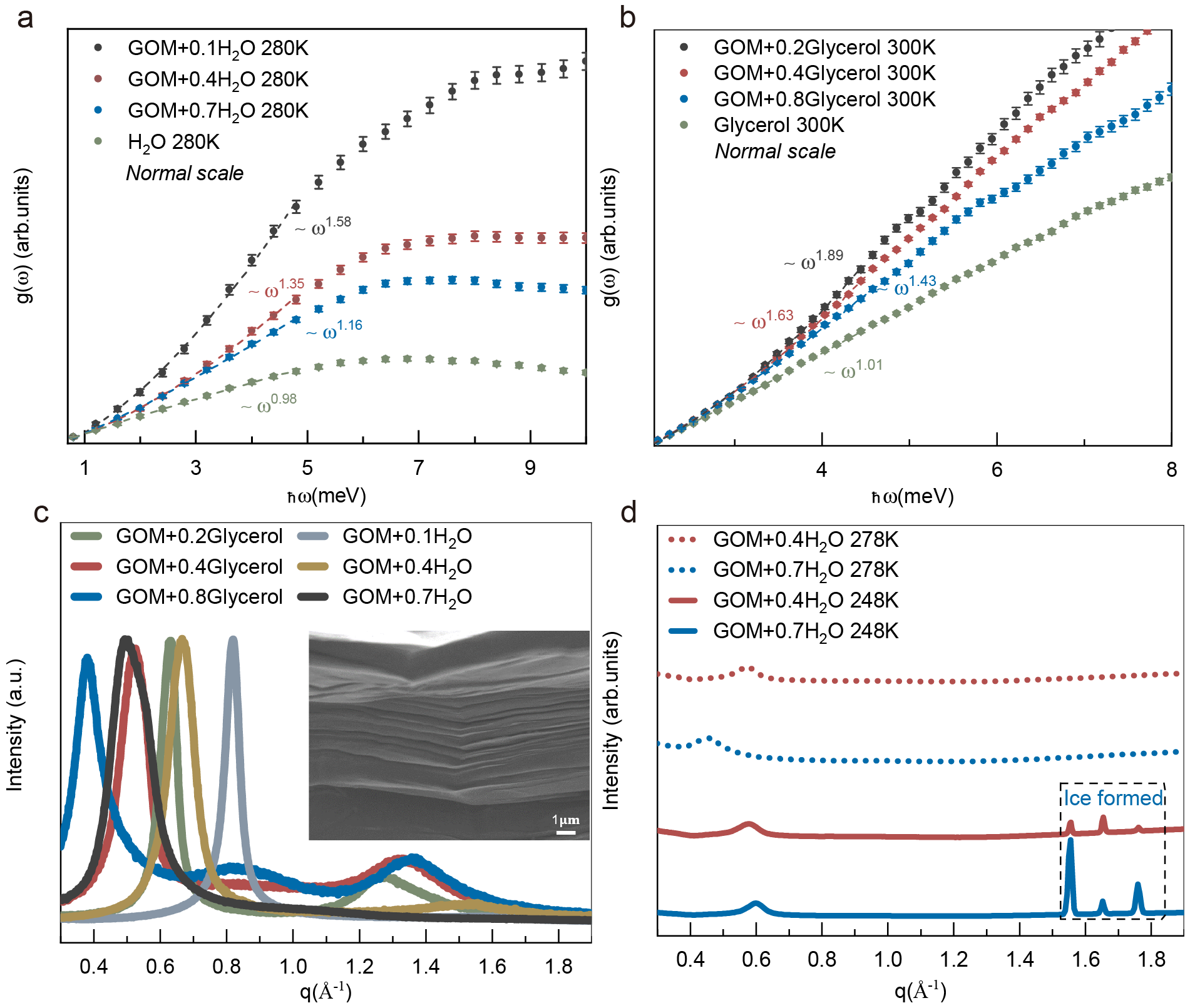}
\caption{\textbf{(a)} The experimental non-normalized \textit{DOS} of confined water at different hydration level (gram water/gram \textit{GOM}). All datasets are fitted in the same energy interval (from $0.8$ meV to $4.5$ meV) and rescaled such that the first data points overlap for better comparison. The error bars are plotted in the same colors as the samples. \textbf{(b)} Analogous experimental non-normalized \textit{DOS} of confined glycerol with different ratio (gram glycerol/gram \textit{GOM}). The error bars are plotted in the same colors as the samples. \textbf{(c)} The XRD data of \textit{GOM} sample with different hydration levels. In the low-$q$ region (below $1\,\angstrom^{-1}$), the sharp and intense characteristic peaks originate from the average interlayer distance of graphene layers in \textit{GOM}, which varies with different liquid contents. For the confined glycerol samples, additional peaks, corresponding to the intra-molecular pairs in glycerol molecules \cite{towey2011structure}, can be observed in the high-$q$ range. The inset is a scanning electron microscope image of \textit{GOM}. \textbf{(d)} The synchrotron X-ray scattering for water confined in \textit{GOM} signaling the absence of Bragg peaks at temperature above the freezing point.}
\label{fig:2}
\end{figure*}

Using inelastic neutron scattering, we measured the \textit{DOS} of water confined within \textit{GOM} at $280$ K and glycerol confined within \textit{GOM} at $300$ K, for various confinement sizes. More details regarding sample preparation and experimental methods are provided in Methods. See also Refs.\citenum{Yu2022,zhang2022conduction} for a similar setup.

As a reference, we present the data for bulk water at $280$ K and glycerol at $300$ K in Figure \ref{fig:2}a  and Figure \ref{fig:2}b  respectively. Both systems exhibit a linear scaling of the \textit{DOS} at low frequencies, consistent with previous experimental results \cite{PhysRevLett.63.2381,dehong2022,jin2024temperaturedependencedensitystates}. We note that the measured energy range of the experimental \textit{DOS} for the confined glycerol begins at larger energy ($\sim 1.8$ meV) than water due to the different observable energy range of the instruments used. Notice also that the results for water are consistent with previous estimates for the structural relaxation time, or equivalently Maxwell relaxation time, that at room temperature are around $1$ ps \cite{PhysRevLett.82.775,MALOMUZH2019111413,PhysRevFluids.4.123302}. This suggests that the dynamics of water at room temperature are likely to exhibit liquid-like behavior (linear in frequency) below approximately $4.13$ meV, in agreement with our analysis.

\begin{table}[!h]  
\begin{tabularx}{\textwidth}{|c|X|X|X|X|X|X|}
\hline
Samples&{\textit{GOM} 0.2gly}&{\textit{GOM} 0.4gly}&{\textit{GOM} 0.8gly}&{\textit{GOM} 0.1H$_2$O}&{\textit{GOM} 0.4H$_2$O}&{\textit{GOM} 0.7H$_2$O}\\ \hline
$2\theta (^\circ)$ & 8.9 & 7.4 & 5.3& 11.4 & 9.2 & 6.9 \\ \hline
Distance$({\angstrom})$ & 10.0 & 12.0 & 16.6 & 7.8 & 9.6 & 12.8\\ 
\hline 
\end{tabularx}
\caption{The interlayer distance for the experimental samples under different hydration levels. The corresponding XRD data for confined water are shown in Figure \ref{fig:2}c.}
\label{table:1} 
\end{table}
These findings are further supported by molecular dynamics simulations (see next section and additional analysis in the Supplementary Information). In stark contrast, under extremely confined conditions, where the water (or glycerol) to \textit{GOM} weight ratio reaches $0.1$ ($0.2$) respectively, the scaling of the low-frequency \textit{DOS} deviates significantly from the linear liquid-like scaling. The \textit{DOS} displays a power law behavior $\omega^\alpha$ where the exponent $\alpha$ lies between the quadratic Debye law for solids and the linear in frequency law for liquids, \textit{i.e.}, $1<\alpha<2$. The relationship between the weight ratio (liquid to \textit{GOM}) and the confinement size is examined by X-ray diffraction in Figure \ref{fig:2}c, and the detailed results are presented in Table \ref{table:1}, where the inter-layer distance in \textit{GOM} varies from 7.8 $\angstrom$ to 13 $\angstrom$ for water when this ratio changes from 0.1 to 0.7. For glycerol, it varies from 10 $\angstrom$ to 16.6 $\angstrom$ when the ratio changes from 0.2 to 0.8. The extreme confinement size provided by \textit{GOM} for both liquids is approximately 7 $\angstrom$ (see Figure \ref{fig:2}c). At that scale, the power $\alpha$ reaches the value of $\approx 1.58$ and $\approx 1.89$ respectively for water and glycerol. By fitting with a power-law function, we find that the scaling of the low-frequency \textit{DOS} of confined water and glycerol gradually deviates from the linear scaling of the bulk liquid as the confinement size is reduced. This suggests that under unidirectional confinement, the influence on low-frequency vibrational modes in liquids positively correlates with the degree of confinement, and the scaling tends to approach the Debye value typical of solids.

\begin{figure}[h!]\centering
\includegraphics[width=\linewidth]{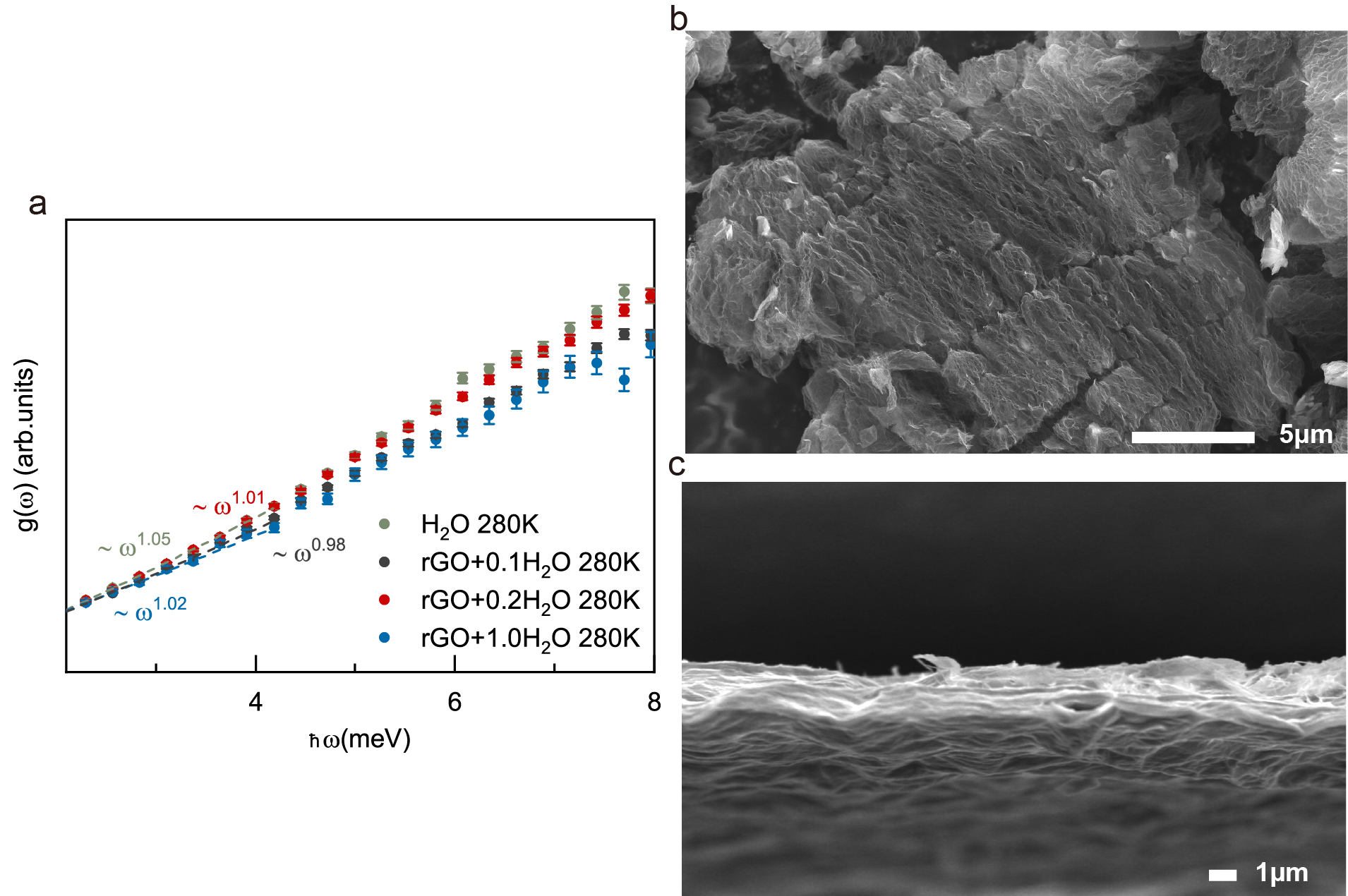}
\caption{\textbf{(a)}The experimental non-normalized \textit{DOS} of water on \textit{rGO} with different hydration levels.\textbf{(b)} The scanning electron microscope image of \textit{rGO}.\textbf{(c)} The scanning electron microscope image of the surface of \textit{GOM}. }
\label{fig:3}
\end{figure}

In our previous work \cite{Yu2022}, we have demonstrated that the \textit{DOS} of \textit{GOM} itself does not contribute significantly to the \textit{DOS} of the confined system in this frequency range. Moreover, we conducted inelastic neutron scattering on water at the surface of reduced graphene oxide (\textit{rGO}) with the absence of any confinement effect. Both \textit{rGO} and \textit{GOM} have rough surfaces due to the presence of oxygen-containing groups, as shown in the scanning electron microscope images in Figure \ref{fig:3}. However, unlike \textit{GOM}, \textit{rGO} does not provide the quasi-two-dimensional confined environment with a sandwich-like structure, which will result in fundamental differences in the spectrum of the confined liquid. As shown in Figure \ref{fig:3}, with only oxygen-containing groups and surface effects \cite{liu2020heterogeneity}, the low-frequency scaling of \textit{DOS} for water exhibits minimal differences. The scaling of the low-frequency \textit{DOS} for three different water contents on \textit{rGO} samples remains approximately linear at temperatures above freezing point. We note that even in the case of the lowest water content (h=0.1), there is no significant change in the scaling compared with pure water. Therefore, the significant changes in the low-frequency \textit{DOS} scaling reported in Figure \ref{fig:2} are primarily due to the quasi-2D geometric confinement provided by \textit{GOM}, rather than merely surface effects. This will be later confirmed by molecular dynamics (\textit{MD}) simulations, where the interactions between the confining material and the confined liquids are for simplicity neglected. Additionally, as shown in Figure \ref{fig:2}d , variable-temperature synchrotron X-ray scattering is employed on \textit{GOM} samples with confined water to demonstrate that the liquid phase is maintained at our experimental temperatures without the formation of a crystalline phase. The presence of Bragg peaks, which symbolize crystalline solid ice, only occurs at lower temperatures, below $270$ K. At temperature above the freezing point, proved by the absence of characteristic Bragg peaks, water confined within \textit{GOM} remains in the liquid state. The diffusive behavior of the water confined in \textit{GOM} has also been reported in our previous work \cite{zhang2022conduction}. Therefore, we conclude that the low-frequency scaling does not originate from a crystallization process but it is rather a property of the confined liquid water itself. 

\begin{figure*}[!t]\centering
\includegraphics[width=\linewidth]{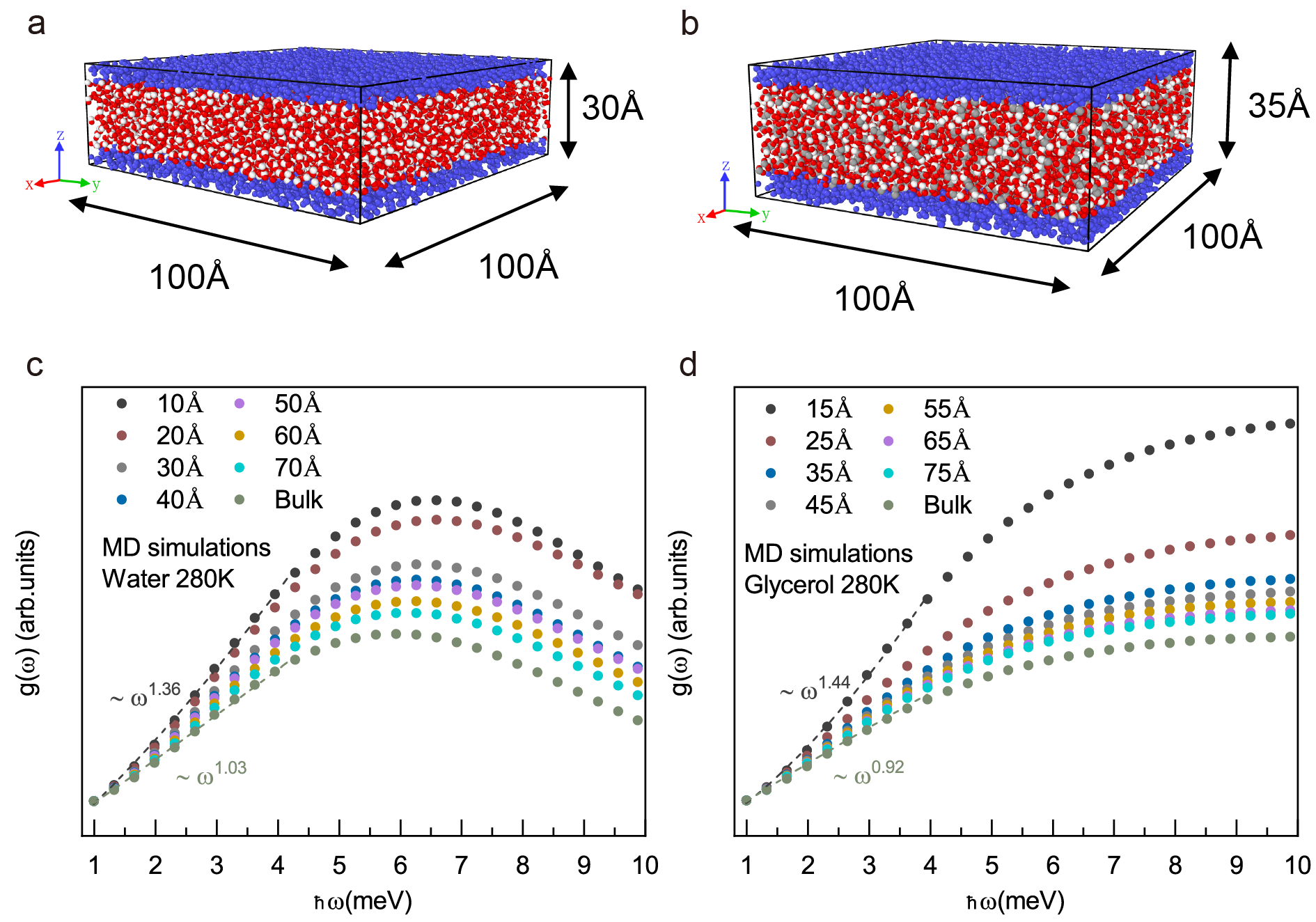}
\caption{The snapshot of the confined system for water \textbf{(a)} and glycerol \textbf{(b)}. \textbf{(c)} The simulated low-frequency \textit{DOS} for confined water under different confinement sizes $L$ (reported in the inset). \textbf{(d)} The simulated low-frequency \textit{DOS} for confined glycerol under different confinement sizes $L$ (reported in the inset). In both panels, only the hydrogen atoms are considered to calculate the \textit{DOS}. The dashed lines indicate the fitted values for the low-frequency scaling at the lowest and highest degree of confinement.}
\label{fig:4}
\end{figure*}

To confirm the experimental results of this faster-than-linear scaling in confined liquids, we conducted molecular dynamics simulations for water and glycerol. A snapshot of the two setups is provided in the top panels of Figure \ref{fig:4}. NPT and NVT ensembles were used to equilibrate and relax the entire system,\cite{perkins2013molecular,yeganegi2012molecular} with the final equilibrium densities of water and glycerol being 0.99 and 1.22 g/cm³, respectively, as previously reported in the literature\cite{abascal2005general,jahn2014effects} (see more details in Methods). The dimensions in the xy plane were set at 100 Å, with the z-direction being used to adjust the confined direction. In the \textit{MD} simulations, we utilized additional confinement sizes with respect to the experiments to validate the experimental results in depth. More details regarding the \textit{MD} simulations can be found in Methods. For the water and glycerol systems, the confinement sizes ranged respectively from $10\,\angstrom$ to $100\,\angstrom$  and $15\,\angstrom$ to $100\,\angstrom$ (see Figures \ref{fig:4}a  and \ref{fig:4}b). In both of the systems, the largest size is considered as a bulk sample. To further verify the impact of the overall dimensions, additional models with $80 \angstrom \times 80 \angstrom$ and $120 \angstrom \times 120 \angstrom$ size of the xy plane were considered. The results are reported in Figure \ref{fig:size} in the Supporting Information. Although there are fluctuations in the numerical values of the low-frequency scaling, the qualitative results remain consistent as the confined dimensions change confirming the validity of our findings.

As shown in Figures \ref{fig:4}c and \ref{fig:4}d, the results of confined water and glycerol from the \textit{MD} simulations are consistent with the experimental observations. In order to be comparable with neutron scattering experiments, only the trajectories of hydrogen atoms are calculated when calculating \textit{DOS}, which has been shown to qualitatively coincide with the results from all atoms in Figure \ref{fig:simsi} in Supporting Information. More precisely, the low-frequency \textit{DOS} exhibits a larger scaling as the confinement size is reduced, strongly deviating from the linear scaling of the bulk state. For the maximum confinement simulated, we observe respectively a power law scaling $\omega^{1.36}$ for water and $\omega^{1.44}$ for glycerol, which are compatible but slightly smaller than the experimental values. This indicates that the simulations can qualitatively reproduce the experimental conditions but they underestimate the confinement strength. 

Interestingly, both in experiments and simulations we find that for the same level of confinement the power law for glycerol is larger than the one of water. This implies that glycerol exhibits a solid-like dynamics on larger scales with respect to water. Assuming that the critical scale for the emergence of solidity is roughly given by $L_c \propto v \tau$, we can, at least qualitatively, explain this observation.  In particular, by following Maxwell's idea and interpreting $\tau$ with the Maxwell relaxation time, the critical length becomes proportional to the viscosity of the liquid. That said, glycerol is more viscous than water, $\approx 1400$ times larger at room temperature, and therefore the solid-like dynamics are expected to emerge at larger scales, as observed in both experiments and simulations.

In the previous section, by means of experiments and simulations, we have shown that by decreasing the confinement size from several nanometers to $7 \angstrom$ the low-frequency behavior of the density of states of water and glycerol drastically changes. In particular, we observed a continuous transition in the \textit{DOS} from a linear scaling in frequency dependence in bulk liquids to a solid-like Debye law in strongly nanoconfined liquids. 

In order to explain this behavior, we rely on the interpretation of liquid dynamics originally proposed by Maxwell and Frenkel \cite{frenkel} (see also Zwanzig \cite{PhysRev.156.190}), and later re-interpreted by Trachenko and collaborators \cite{trachenko2015collective,BAGGIOLI20201}. Mathematically, this framework can be described using the telegrapher equation, Eq.\eqref{kgapeq}, which determines the dynamics of the collective (shear) modes in liquids, up to the lowest order in the frequency $\omega$ and the wave-vector $k$ and provides a simple interpolation between a solid-like behavior at high frequencies and a liquid-like dynamics at low frequencies.

By solving Eq.\eqref{kgapeq}, and focusing on the real part of the frequency, one finds the simple expression:
\begin{equation}
    \mathrm{Re} (\omega) = v \sqrt{k^2-k_g^2}\qquad \text{where}\qquad k_g\equiv \left(2 v \tau_{g}\right)^{-1},\label{ee}
\end{equation}
which is valid in the vicinity of the $k$-gap, but not at very large $k$. Here, we have kept a distinction between the structural relaxation time $\tau$ ($\approx 1$ps for bulk water) and the timescale $\tau_g$ appearing in the k-gap equation above.
In bulk liquids, $\tau_g$ decreases with temperature, and therefore $k_g$ can be taken as a measure of ``fluidity''. 

Using \textit{MD} simulations, we obtained the dispersion relation of the collective shear waves in liquid water, as shown in Figure \ref{fig:4}a. Specifically, we calculated the current-current correlation function of transverse waves, $C_T(q,\omega)$, and extracted its maxima to ascertain the complete dispersion relation for the transverse waves\cite{kryuchkov2019excitation,PhysRevLett.118.215502}. Unfortunately, the statistical quality of the data points and the fitting curve in the low-wavenumber region is poor. This limitation is attributed to the finite size of our simulations ($100\angstrom$). For the bulk sample at $280$ K, we find that $k_g \approx 0.23\, \angstrom^{-1}$. which is of the same order as liquid gallium at $313$ K \cite{PhysRevB.101.214312} and liquid sodium at $393$ K \cite{doi:10.1073/pnas.1006319107}. Most importantly, the obtained value is in quantitative agreement with a previous estimate for water at room temperature, where the system has been proven to be fully hydrodynamic below a value of $\approx 0.2 \,\angstrom^{-1}$ \cite{PhysRevLett.79.1678}.

It should be noted that, as derived in Ref.\citenum{PhysRevLett.79.1678}, the dynamics of bulk water at room temperature becomes underdamped as is the case in solids only above $\approx 0.4 \,\angstrom^{-1}$. This implies that the requirement of having the wave-vector larger than the k-gap $k_g$ is not enough to have a well-defined underdamped solid-like dynamics in liquids. As explained in Supporting Information, this is consistent with k-gap theory. More precisely, above but near the k-gap, $\mathrm{Re}(\omega)<\mathrm{Im}(\omega)$, meaning that the dynamics are not liquid-like but still overdamped. This is consistent with the range of $0.2\, \angstrom^{-1}<k<0.4 \,\angstrom^{-1}$ found experimentally in Ref.\citenum{PhysRevLett.79.1678}. More interestingly, at a quantitative level, $0.4 \,\angstrom^{-1}$ is approximately twice of the k-gap, and converted in frequency scale is very close to the Frenkel criterion (see Supporting Information for more details about this).

\begin{figure*}[!t]\centering
\includegraphics[width=\linewidth]{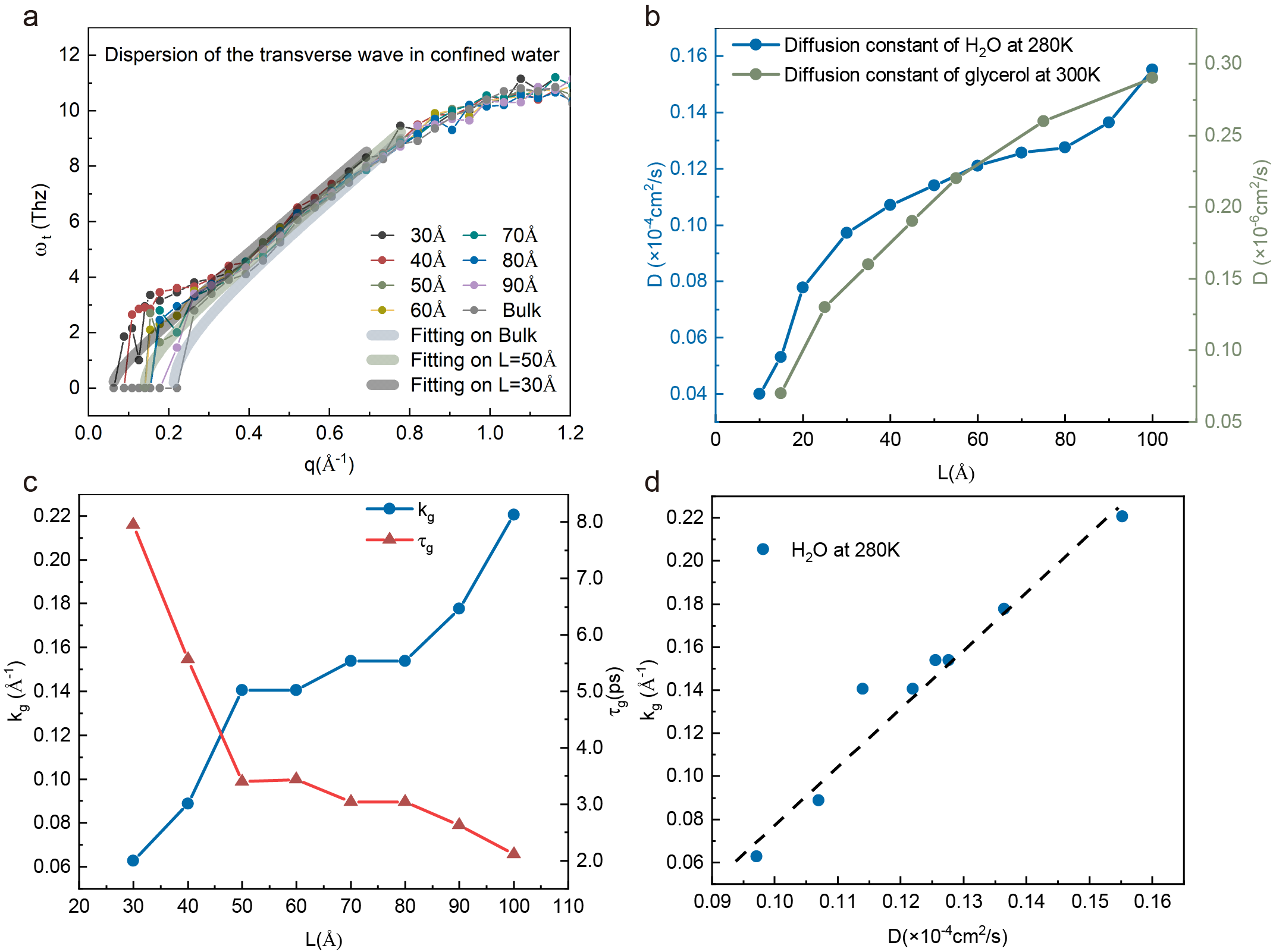}
\caption{\textbf{(a)} Dispersion curves for transverse waves in confined water under different confinement sizes. \textbf{(b)} Self-diffusion constant increasing with the confinement size in confined water and confined glycerol. \textbf{(c)} The momentum gap extracted from the dispersion curves in panel \textbf{(a)} and the corresponding relaxation time $\tau_g$ as a function of the confinement size $L$. \textbf{(d)} The linear correlation between $k_g$ and the diffusion constant.}
\label{fig:5}
\end{figure*}

In order to prove the emergence of solidity under confinement, in Figure \ref{fig:5}a, we analyze the dispersion relation of the collective shear waves in the water sample by reducing the confinement size. The detailed results of the fit can be found in Table \ref{table:2} in Supporting Information. We observe that the momentum gap of the collective modes decreases with decreasing confinement size. In other words, under strong confinement the gap in the wave-vector closes and the dispersion relation shifts towards the solid-like one $\mathrm{Re}(\omega)= v k$. Before analyzing this behavior in detail, we seek further confirmation of this trend by examining the experimental and simulation diffusion constants as a function of the confinement size. 

As shown in Figure \ref{fig:5}b, the diffusion constant obtained from the mean square displacement (see Figure \ref{fig:e3}) increases monotonically with the confinement size $L$ and approaches the value for bulk water for larger $L$. A decrease of the self-diffusion constant with confinements size, and a dramatic increase of the viscous forces have been already reported in simulations and experiments \cite{zhang2022conduction,PhysRevLett.100.106102,KUO2023140612,Ortiz-Young2013}.
For comparison, in the same panel, we also show the behavior of the diffusion constant for confined glycerol at $300$ K. The diffusion constant of glycerol is approximately two orders of magnitude smaller than that of water, as the dynamics therein are much slower and the effective viscosity is much larger.

We then extract the parameters $v$ and $\tau_g$. Despite the precision of the fitting is limited, specially for very low values of $L$, the analysis can still provide a useful qualitative estimate. In Figure \ref{fig:5}c, we show the behavior of $k_g$ and $\tau_g$ as a function of the confinement size $L$. As explicitly verified in Supporting Information (see Table \ref{table:2}), the shear waves speed (and the speed of longitudinal sound as well) are to a first approximation constant with respect to $L$. On the contrary, the relaxation time $\tau_g$, and consequently the momentum gap $k_g$, are very sensitive to the confinement scale. The relaxation time $\tau_g$ grows with $L$, and the momentum gap follows approximately the inverse behavior $k_g\propto 1/\tau_g$. From a physical perspective, these results indicate that, by increasing the strength of the confinement, collective shear waves in the liquid start to propagate at lower frequencies and the solid-like dynamics extend to lower wave-vectors. Moreover, the dynamics and the local-rearrangements are substantially slowed down, as a further evidence that the liquidity of the system is hindered at short distances. 

We notice that for the bulk water the relaxation time $\tau_g$ is $\approx 2$ps. This is about a factor of two larger than the structural relaxation time and the Maxwell relaxation time reported for bulk water at room temperature \cite{PhysRevLett.82.775,MALOMUZH2019111413,PhysRevFluids.4.123302}. As a final remark, the k-gap analysis and the behavior of the diffusion constant confirm a close connection between the retardation of relaxation dynamics and the tendency toward solid-like vibrational motion. The same conclusion has been reached in several \textit{MD} simulations by changing the temperature of the system \cite{khusnutdinoff2020collective,PhysRevLett.118.215502}. 

In Figure \ref{fig:5}d, we reveal a direct linear correlation between the size of the momentum gap for collective shear waves and the single-particle self-diffusion constant. {The two quantities show an approximate linear relation of the type $k_g \propto D$ confirming that they are both measurements of ``fluidity''. The larger the gap of shear waves, the larger the self-diffusion constant and the ability of the system to flow as a fluid. To the best of our knowledge, this direct correlation between the k-gap and the diffusion constant has not been revealed before, and it certainly deserves further investigation.

At this point, it is reasonable to ask how the transition from the linear in frequency liquid-like behavior to the quadratic Debye law happens upon dialing the confinement size $L$ and how it is related to the re-appearance of collective shear waves and the behavior of the momentum gap $k_g$. In order to reach a qualitative argument, a simplified toy model based on the k-gap equation Eq.\eqref{kgapeq} can be constructed (see Supporting Information for more details). By neglecting the effects of the other excitations and concentrating on the gapped shear waves, and by assuming as a first order approximation a spherical phase space for their wave-vector unaffected by the confinement size, one can derive a simple expression for the contribution of the shear waves to the \textit{DOS} given by:
\begin{equation}\label{ddd}
    g_{\text{shear}}(\omega)\propto \omega \sqrt{\omega^2+\frac{1}{4 \tau_g^2}}\,.
\end{equation}
Despite the simplicity of this derivation, and its many assumptions, one striking prediction emerges. Precisely, a transition from a liquid-like linear \textit{DOS} to a Debye law is expected at frequencies $\omega \gg 1/\tau_g$, that is another manifestation of the Frenkel criterion. In other words, we expect that the deviations from the purely linear behavior will appear only at frequencies larger than the inverse relaxation time $\tau_g^{-1}$. Since $\tau_g$ increases by reducing the confinement size $L$, as shown in Figure \ref{fig:4}c, one would expect the Debye scaling to reach lower and lower frequencies upon confining the liquid. 

In order to check this prediction, in Figure \ref{fig:6}a, we revisit the experimental data for confined water under this perspective and we estimate the frequency at which the linear behavior is lost as a function of the confinement size $L$. We denote such an energy scale as the crossover frequency $\omega_\times$. {In bulk water at $280$ K we do not observe any crossover and the low-frequency \textit{DOS} is perfectly linear below $4$ meV. This is in perfect agreement with the idea that the solid-like dynamics can appear only above $1/\tau_g$. For bulk water, $\tau_g=1$ ps and therefore the cutoff for solid-like vibrations is around $4.15$ meV. 

As proved in Figure \ref{fig:5}, under confinement the relaxation time strongly increases, moving the onset for solid-like vibrations in confined liquids to energies below the bulk cutoff of $4$ meV. Indeed, for higher degrees of confinement -- shorter $L$ -- we start observing a crossover between a linear behavior, that appears within the blue region in Figure \ref{fig:6}a, and a quadratic Debye-like one that emerges above $\omega_\times$, in the white region in Figure \ref{fig:6}a. These two different scalings are emphasized in Fig.\ref{fig:6}a by dashed black lines. The crossover frequency becomes smaller by decreasing the confinement size $L$, ranging from $\approx 2.8$ meV at $L=12.8 \,\angstrom$ to $1.9$ meV at $L=7.8 \,\angstrom$. 

This is compatible with the relaxation time becoming longer and longer under the enhanced confinement and the onset of solid-like dynamics, $\sim 1/\tau_g$, appearing at lower frequencies, below the experimental cutoff of $\approx 4$ meV (see Figure \ref{fig:6}b  for a representation of this phenomenon).\\
This simple argument also suggests that the fractional power laws extracted in the previous section from the experimental data using a single power fitting function might be the result of a combination of a liquid-like linear term and a Debye quadratic contribution. Qualitatively, the smaller the confinement size $L$, the more the Debye contribution enters at low frequency and the larger the power extracted from a single power fitting law. This also explains why a perfect Debye scaling is not recovered for small values of $L$. This would happen only in the unrealistic limit $\tau_g \rightarrow \infty$ in which the crossover frequency goes to zero and the relaxational dynamics are completely frozen. For the shortest confinement scale we can achieve, $L \approx 7.8\, \angstrom$, we see a crossover frequency of $\omega_\times \approx 1.9$ meV, implying that below such a scale a residual linear in frequency liquid-like contribution is still present. \\

\begin{figure}\centering
\includegraphics[width=\linewidth]{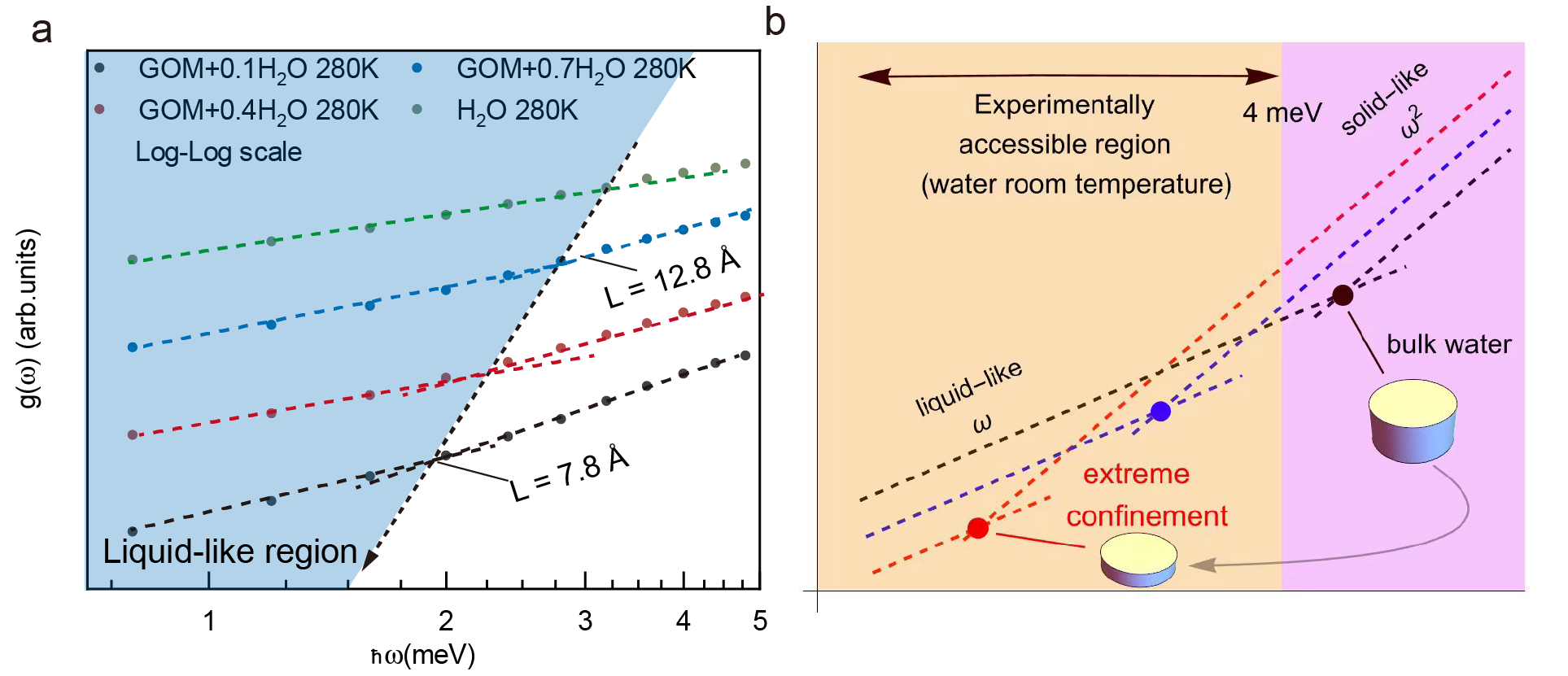}
\caption{\textbf{(a)} A re-visitation of the experimental data in Figure \ref{fig:2} using 
a log-log scale. The colored region represents the liquid-like regime, where solid-like vibrational modes cannot exist. The black dashed lines indicate the two different scalings $g(\omega)\sim \omega$ (in the blue region) and $g(\omega)\sim \omega^2$, above $\omega_\times$ in the white region. \textbf{(b)} A pictorial representation of the underlying physics based on k-gap theory (see Supporting Information). The lines from black to red corresponds to smaller confinement size $L$. The colored disks locates the position of the crossover between linear behavior and quadratic behavior -- the onset of solid-dynamics -- upon confinement. The orange region indicates the low-frequency regime, where the power-law scalings can be experimentally detected, which for water at room temperature extends up to $\approx 4$meV. For bulk water at room temperature, solid-like vibrations are expected only above that scale, within the magenta region.}
\label{fig:6}
\end{figure}

\subsection{Conclusion}
In this study, we investigated the density of states of water and glycerol liquids nano-confined in multiple-layer graphene oxide membranes using inelastic neutron scattering and molecular dynamics simulations. We focused on the low-frequency (between $1$ meV and $4$ meV) scaling of the \textit{DOS} in confined liquids, its variation with the confinement length and its relation with the relaxational dynamics and the dispersion of gapped shear waves. Experimental observations revealed that the low-frequency \textit{DOS} of the confined liquid evolves from a linear behavior in the bulk state to a faster scaling, close to the Debye-like quadratic behavior, under strong confinement. We note that based on other confined liquids and confining materials, the varying affinities between the two may yield different results, which represents a promising direction for future exploration.

Through molecular dynamics simulations, we explored a broad range of confinement sizes and confirmed the gradual solid-like evolution of both liquids as the confinement size decreases. Analyzing the vibrational modes in the different directions, we found that emergent solid-like vibrational modes primarily arise along the confined direction, in agreement with previous simulation results \cite{KUO2023140612}, and that spatial confinement strongly affects the dispersion relation of transverse waves in the liquid. As the confinement size decreases, both the diffusive processes and the structural rearrangements are slowed down, and the transverse dispersion curves progressively resemble those of a solid, with no wave-vector gap. Upon confinement, the longitudinal waves of the liquid are not significantly affected. On the contrary, the momentum gap for collective shear waves, typical of liquids, closes by decreasing $L$ and beautifully correlates with the behavior of the self-diffusion constant $D$. In a nutshell, decreasing the confinement size $L$ is akin to decreasing the temperature $T$ and going towards the solid phase in a continuous fashion. In addition, we have found that the relaxation time $\tau_g$ extracted from the k-gap equation is strongly slowed down upon confinement.

Building upon Frenkel-Maxwell's ideas \cite{frenkel} and the more recent k-gap theory \cite{BAGGIOLI20201}, we provided a simple physical picture for the experimental and simulation observations. In particular, we have corroborated that the appearance of solid-like dynamics at low frequency is highly correlated with slow down of the relaxation time under confinement. This confirms the idea that at frequencies above the inverse of the relaxation time $1/\tau$, solid-like vibrational modes appear. Under nanoconfinement conditions, the corresponding energy scale becomes smaller than $4$ meV and strongly alters the low-frequency \textit{DOS} of liquids. This mechanism is accompanied by the reduction of the gap for transverse shear waves, which is again controlled, according to k-gap theory, by the inverse of the relaxation scale. In other words, we have shown that the closing of the k-gap under confinement is the microscopic origin behind the solid-like low-frequency \textit{DOS} of liquids under confinement and it strogly correlates with the decrease of the self-diffusion constant. This whole picture supports Frenkel's idea of continuity between the liquid and solid phases of matter \cite{Frenkel1935}. We also notice that our work primarily focuses on the low-frequency information of the \textit{DOS} in confined liquids, aiming to elucidate the scaling changes and connections between liquid-like and solid-like \textit{DOS}. The high-frequency vibrational information of confined liquids and solids also presents an attractive prospect for future research, as extensively explored using infrared (IR) or Raman spectroscopy in existing studies.\cite{mallamace2007evidence,stefanutti2019vibrational,erko2011confinement} In conclusion, our study not only bridges theoretical physics with applied nanotechnology but also underscores the potential of nanoconfinement in modulating liquid properties for advanced material science applications.

\section{Methods}
\textbf{Sample preparation}
The \textit{GOM} sample was synthesized using the modified Hummers' method\cite{marcano2010improved}. Initially, the sample was dehydrated by heating it from room temperature to 313 K and subsequently annealed at this temperature for 12 hours under vacuum conditions to achieve a dry state. The oxidation rate of the \textit{GOM} sample was determined to be $28\%$ using X-ray Photoelectron Spectroscopy (\textit{XPS}).
After dehydration, the \textit{GOM} sample was sealed in a desiccator and exposed to water vapor to facilitate the adsorption of water molecules onto the surface and between the interlayers of the \textit{GOM} sheets. The exposure time of the sample was adjusted to control the hydration levels. The final hydration levels were determined by measuring the weight of the sample before and after water adsorption, providing a quantitative assessment of the absorbed water content. The \textit{rGO} powder sample was purchased from Nanjing XFNANO Materials Tech Co., Ltd., China. The subsequent procedures for water adsorption were consistent with \textit{GOM}. The dehydrated \textit{GOM} is soaked in glycerol for varying amounts of time and dried at 40$^{\circ}$C for 24 hours after removal to thoroughly remove the water. The specific gravity of glycerol is determined by the soaking time of the samples.\\
\textbf{Powder X-ray Diffraction (\textit{PXRD})}
The powder X-ray diffraction (\textit{PXRD}) data for \textit{GOM} at various hydration levels were obtained using a Rigaku Mini Flex600 X-ray diffractometer. The instrument was equipped with a Cu K$\alpha$ source ($\lambda$ = 1.5406 $\angstrom$) and operated at 40 kV and 15 mA. The data were collected over a scanning range of 10$\degree$ to 60$\degree$ at a scan rate of 10$\degree$/min. The analysis of the \textit{PXRD} data was performed using MDI Jade software.\\
\textbf{Synchrotron X-ray scattering }
Synchrotron X-ray scattering measurements were utilized to track the changes in interlayer distance within graphene-oxide membranes during temperature reduction and ice formation. The experiments were conducted at the BL16B1 beamline of the Shanghai Synchrotron Radiation Facility (\textit{SSRF}) using X-rays with a wavelength of $1.24 \angstrom$. The scattering patterns were captured using a Pilatus 2M detector, featuring a resolution of 1475 pixels $\times$ 1679 pixels and a pixel size of $172 \,\mu m \times 172\, \mu m$. To ensure accurate data collection, each frame had an acquisition time of 10 s. The sample-to-detector distance for measurements was maintained at 258 mm.\\
\textbf{Scanning Electron Spectroscopy (\textit{SEM})}
The SEM images were acquired using a MIRA 3 FE-SEM operating at a 5 kV accelerating voltage.\\
\textbf{Inelastic Neutron Scattering (\textit{INS})}
Due to the significantly larger incoherent scattering cross section of hydrogen atom\cite{zheng2022universal}, the intensity measured in neutron signals in the system are primarily influenced by the incoherent scattering function, which predominantly reflects the self-motion of water molecules within the sample (the contribution from the graphene oxide will be very small\cite{Yu2022}). The experimental density of states (\textit{DOS}), denoted as $g(\omega)$, can be derived from the dynamic structure factor, $S(q,\omega)$, using the approach\cite{1997connection}:
\begin{align}\label{expexp}
    g(\omega)\, = \,\int \frac{\hslash\omega}{{q}^2}S(q,\omega)\left(1-e^{-\frac{\hslash\omega}{k_B T}}\right)dq,
\end{align}
where $ \hslash $ is the Planck constant, $q$ is the scattering wavevector, $\omega$ is the frequency which related to the energy transfer, $k_B$ is the Boltzmann constant, and T is the temperature.

The experimental \textit{DOS} obtained from Eq. \eqref{expexp} is the generalized \textit{DOS} which is a neutron- weighted result over all elements contained in the sample, though it may be dominated by hydrogen atoms.  Multiple scattering is minimized by choosing sample thickness such that about $10\%$ scattering is ensured. Multi-phonon scattering effect is expected to be small at the low energy range considered here.

For water confined in \textit{GOM}, three samples were fabricated with different weight ratio \textit{h} (gram water/gram \textit{GOM}) : 0.1, 0.4 and 0.7. The inelastic neutron scattering experiment were conducted by the time-of-flight cold neutron spectrometer - PELICAN at Australian Nuclear Science and Technology Organisation (\textit{ANSTO}). The incident neutron energy is 14.9 meV with an energy resolution $\Delta E = 0.5$ meV at the elastic peak\cite{pelican}. The $q$ range covered is from $0.08 \,\angstrom ^{-1}$ to $4.5 \,\angstrom ^{-1}$.
The samples were contained inside aluminum foils which were  sealed in aluminum sample cans under helium atmosphere. The empty can signal was subtracted as background at each temperature. The detector efficiency was normalized using a vanadium standard. Data reduction and \textit{DOS} extraction were performed by LAMP software package\cite{lamp} and the scripts are available upon request.
The experiments for glycerol confined in \textit{GOM} and water on \textit{rGO} were conducted using the cold neutron multi-chopper spectrometer LET at ISIS in UK. The confined glycerol samples were also fabricated in three different weight ratio \textit{h} (gram glycerol/gram \textit{GOM}): 0.2,0.4 and 0.8 The measurement was done with incident energy $22.78$ meV which covers the energy transfer range up to $13.5$ meV and the $q$ range from $0.34 \,\angstrom^{-1}$ to $7.08 \,\angstrom^{-1}$. 
Data reduction was performed on Mantid software packages\cite{mantid}.\\
\textbf{Molecular dynamics (\textit{MD}) simulations}
The classical \textit{MD} simulation was performed via the large-scale atomic/molecular massively parallel simulator (LAMMPS)\cite{lammps} to simulate water and glycerol at $280$ K and $300$ K, respectively. For the bulk samples, the total number of the atoms are respectively 99603 and 98000 for water and glycerol. Water molecules are 
simulated with the widely used four-point TIP4P/2005\cite{tip4p2005} model where the energy parameter $\epsilon_{O-O}$ is $0.008$ eV 
and the size parameter $ \sigma_{O-O}$ is $3.16$ \angstrom.  
The interactions potential of water molecules is modeled by 12-6 Lennard-Jones (LJ) potential with a cutoff radius of $12$ $\angstrom$  and the electrostatic forces are calculated with PPPM algorithm with an accuracy of $10 ^{-4}$. The system of glycerol was modeled using Material Studio and simulated using the AMBER force field parameters, which have been proven to accurately describe the dynamics and structure of the glycerol system \cite{dosglycerol}. All initial structures were relaxed at given temperatures and a pressure of 1 atm  with the isothermal-isobaric (NPT) ensemble for 300 ps. The Nos\'e-Hoover thermostat and Parrinello-Rahman barostat to control the temperature and pressure, allowing the whole system to reach thermodynamic equilibrium at predefined temperature and pressure conditions after relaxation.\cite{nose1984unified,hoover1985canonical} The systems were then switched to NVT ensemble and allowed to relax for 100 ps, during which the number of particles, temperature, and volume were kept constant to calculate the dynamic properties. The Newton equation of motion was integrated using the velocity-Verlet algorithm with a time step of $1$ fs per step. We freeze the bottom and top layer ($\sim3$  \angstrom) of the slab systems to force the structure to remain 2D confined during the whole simulation.

A snapshot of the slab geometry can be found in Figure ~\ref{fig:3} a,b in the main text.
\noindent The \textit{DOS} is calculated by the Fourier transform of the velocity auto-correlation function:
\begin{equation}
C_v(\omega)=\int_{-\infty}^\infty C_v(t)\exp(-i\omega t)dt\,.
\end{equation}
The velocity auto-correlation function (\textit{VACF}) is defined as:
\begin{equation}
C_v(t)=<\textbf{v}(0)\cdot \textbf{v}(t)>
\end{equation}
where $\textbf{v}(0)$ are the oxygen velocities. In order to compare with the \textit{DOS} from neutron experiments, which mainly comes from the signal of hydrogen atoms, we calculated the \textit{VACF} with the corresponding \textit{DOS} of hydrogen atoms and all atoms(see Figure \ref{fig:3} and Figure \ref{fig:simsi}), respectively. \\
The Diffusion constant is calculated by the mean square displacements of the water molecules:
\begin{align}
D = \frac{1}{6}\frac{d}{dt}\left<\left|\mathbf{r}_i(t)-\mathbf{r}_i(0)\right|^2\right>,
\end{align}
where $\mathbf{r}_i(t)$ is the position of the $i^{th}$ particle at time $t$.\\

The current density is a vector quantity which can be decomposed
into a longitudinal part containing the component parallel to the
$\boldsymbol{q}$-vector and a transverse part containing the perpendicular component, according to
\begin{align}
\boldsymbol{j}(\boldsymbol{q},t)=\boldsymbol{j}_L(\boldsymbol{q},t)+ \boldsymbol{j}_T(\boldsymbol{q},t)
\end{align}
where we have defined
\begin{align}
&\boldsymbol{j}_T(\boldsymbol{q},t)=\sum_i^{N}(\boldsymbol{v}_i(t)\cdot\hat{\boldsymbol{q}})\hat{\boldsymbol{q}}e^{i\boldsymbol{q}\cdot \boldsymbol{r}_i(t)},\\
&\boldsymbol{j}_T(\boldsymbol{q},t)=\sum_i^{N}[(\boldsymbol{v}_i(t)-(\boldsymbol{v}_i(t)\cdot\hat{\boldsymbol{q}})\hat{\boldsymbol{q}}]e^{i\boldsymbol{q}\cdot \boldsymbol{r}_i(t)},
\end{align}
and use $\hat{\boldsymbol{q}}$ to denote the unit vector.\\
The current correlation functions can now be computed as:
\begin{align}
C_L(\boldsymbol{q},t)=\frac{1}{N}\left\langle(\boldsymbol{j}_L(\boldsymbol{q},t)\cdot\boldsymbol{j}_L(\boldsymbol{-q},0)\right\rangle    
\end{align}
\begin{align}
C_T(\boldsymbol{q},t)=\frac{1}{N}\left\langle(\boldsymbol{j}_T(\boldsymbol{q},t)\cdot\boldsymbol{j}_T(\boldsymbol{-q},0)\right\rangle    
\end{align}
Then the velocity current spectra could be derived using:
\begin{align}
C_{L,T}(\boldsymbol{q},\omega)=\int dt e^{i\omega t} Re(C_{L,T}(\boldsymbol{q},t)) .  
\end{align}
Since simple fluids are isotropic, we average $C_{L,T}(\boldsymbol{q},\omega)$ over all directions of the wavevector $\boldsymbol{q}$. 
\begin{align}
C_{L,T}(q,\omega)=\frac{1}{N_q}\sum_{|\boldsymbol{q}|=q}C_{L,T}(\boldsymbol{q},\omega)\,,
\end{align}
where $N_q$ is the number of directions used for averaging.

\begin{acknowledgement}

We thank H.~Xu, A.~Zaccone, K.~Trachenko, L.~Noirez and T.~Keyes for fruitful discussions on the topic of liquids. We thank Dr.~Victoria Garcia Sakai from ISIS Neutron and Muon Facility for assistance with the inelastic neutron scattering. We acknowledge the  Instrumental Analysis Center of Shanghai Jiao Tong University for assistance with structural characterization via SEM and XRD. We thank Dr. Xiaran Miao from BL16B1 beamline of Shanghai Synchrotron Radiation Facility (SSRF) for help with synchrotron X-ray measurements. 
This work was supported by the National Natural Science Foundation of China (11974239), the Innovation Program of Shanghai Municipal Education Commission(2019-01-07-00-02-E00076), and Shanghai Jiao Tong University Scientific and Technological Innovation Funds (21X010200843). M.B. acknowledges the support of the Shanghai Municipal Science and Technology Major Project (Grant No.2019SHZDZX01) and the sponsorship from the Yangyang Development Fund. We acknowledge the support of the Australian Centre for Neutron Scattering, ANSTO and the Australian Government through the National Collaborative Research Infrastructure Strategy, in supporting the neutron research infrastructure used in this work via ACNS proposal P7273.

\end{acknowledgement}

\begin{suppinfo}

The following file is available free of charge.
\begin{itemize}
  \item Supporting Information: 
  Further analysis of simulation data: partial density of states and self-diffusion constant; Detailed fitting and theory model: collective waves and k-gap model; Supporting Figures S1--S7; Supporting Table S1.

\end{itemize}

\end{suppinfo}
\section{Associated Content}
\subsection*{Preprint Version}
Yuanxi Yu; Sha Jin; Xue Fan; etc. Emergence of solid-like Debye scaling in the vibrational density of states of liquids under nanoconfinement. 2023,2307.11429. arXiv. \url{https://arxiv.org/abs/2307.11429}

\section{Supporting Information}
\subsection{Additional dynamical properties in MD simulations}
In Figure \ref{fig:e2}, we present the simulation density of states of water under different confinement conditions and decomposed into the three different spatial directions x,y,z. Here, the x, y, and z coordinates are consistent with the labels used in Figure \ref{fig:3}, where x and y define an unconfined two-dimensional plane, and the z-direction is perpendicular to the confining boundaries.
At the smallest confinement size ($L=10\, \angstrom$), the scaling of the density of states (DOS) in the z-direction, which represents the confined direction, is significantly separated from the scaling of the other two components. It grows faster  compared to the horizontal components and its low-frequency power-law is closer to the Debye value. In other words, and perhaps not surprisingly, the the solid-like dynamics is more pronounced along the confined direction. This is in agreement with the results recently appeared in Ref.\cite{KUO2023140612}. The overall DOS exhibits a scaling that falls between the faster one along the $z$ direction and the on-plane one. This might also explains why the full DOS does not exhibit a quadratic Debye law even at the highest degree of confinement. This is simply because the dynamics along the non-confined directions are still mostly liquid-like. This phenomenon becomes less pronounced as the confinement size increases. In the reference case of bulk water, where the anisotropy is lost, the DOS scaling for all three components and the overall DOS perfectly overlap.
\begin{figure}[!t]
    \centering
    \includegraphics[width=\linewidth]{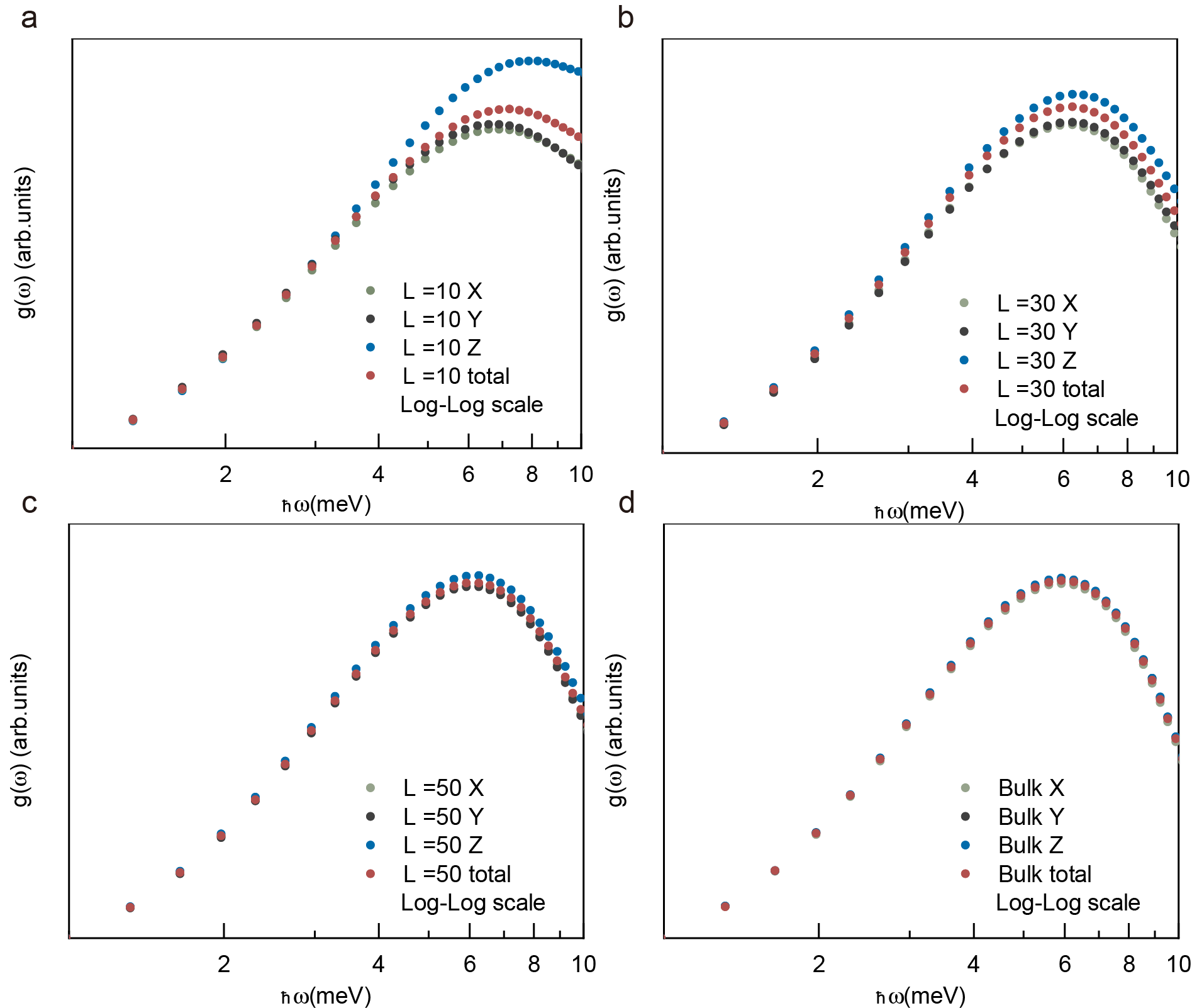}
    \caption{The density of states (DOS) of confined water along different directions and for different confinement sizes: \textbf{(a)} 10 $\angstrom$; \textbf{(b)} 30 $\angstrom$; \textbf{(c)} 50 $\angstrom$; \textbf{(d)} bulk water as reference. Double logarithmic scale is used to highlight the differences in the low energy scalings. The data are vertically shifted for better visualization.}
    \label{fig:e2}
\end{figure}
Finally, in Figure \ref{fig:e3}, we report the behavior of the mean-square displacement (MSD) as a function of time and for different degrees of confinement. By fitting the late-time linear behavior we can directly obtain the self-diffusion constant of the system for various confinement strengths. For both confined water and glycerol, the diffusion constant decreases with confinement as reported in the main text in Figure \ref{fig:5}.

\begin{figure}
    \centering
    \includegraphics[width=\linewidth]{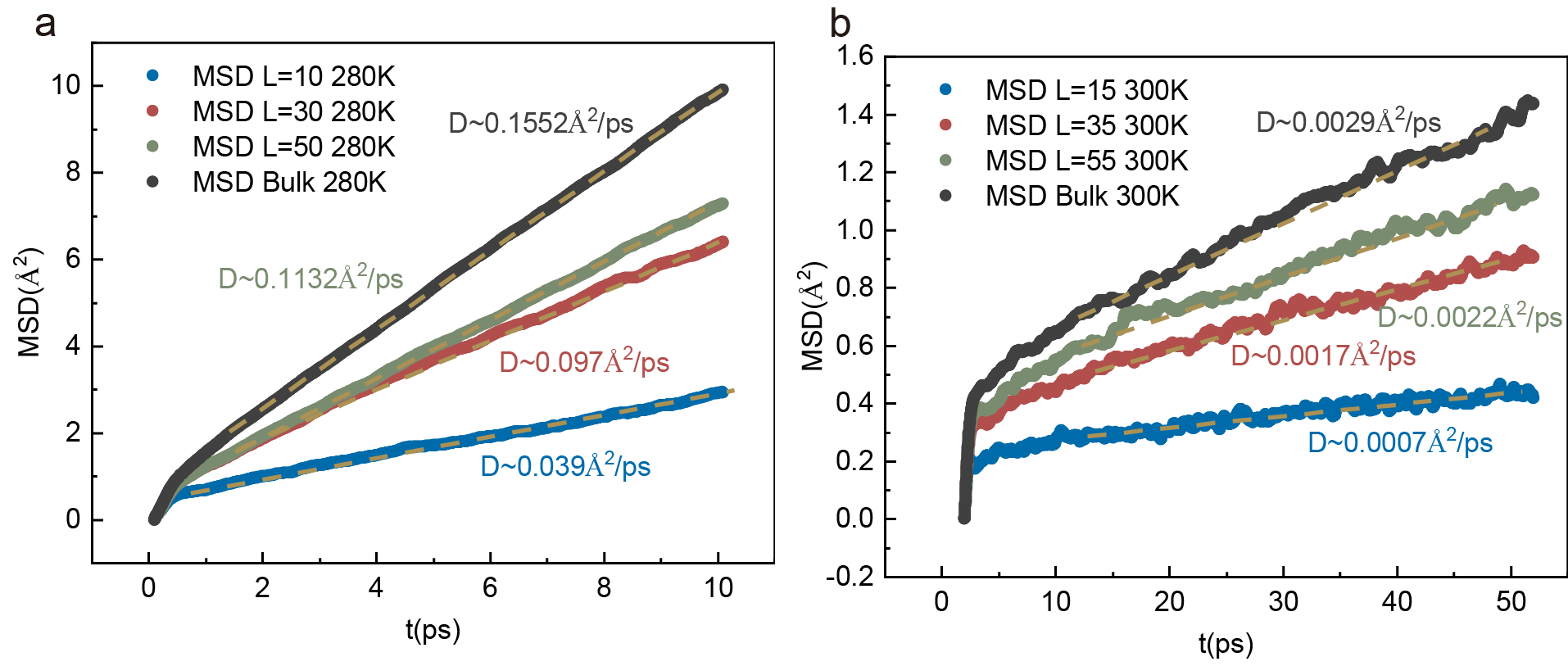}
    \caption{The mean-square displacement for confined water \textbf{(a)} and confined glycerol \textbf{(b)} as a function of time and different degrees of confinement. The dashed lines indicate the late-time linear fit which allows to determine the value of the self-diffusion constant reported as inset. The obtained data for the self-diffusion constant as a function of the confinement size $L$ are reported in the main text in Figure \ref{fig:5}.}
    \label{fig:e3}
\end{figure}

\begin{figure}[h!]\centering
\includegraphics[width=\linewidth]{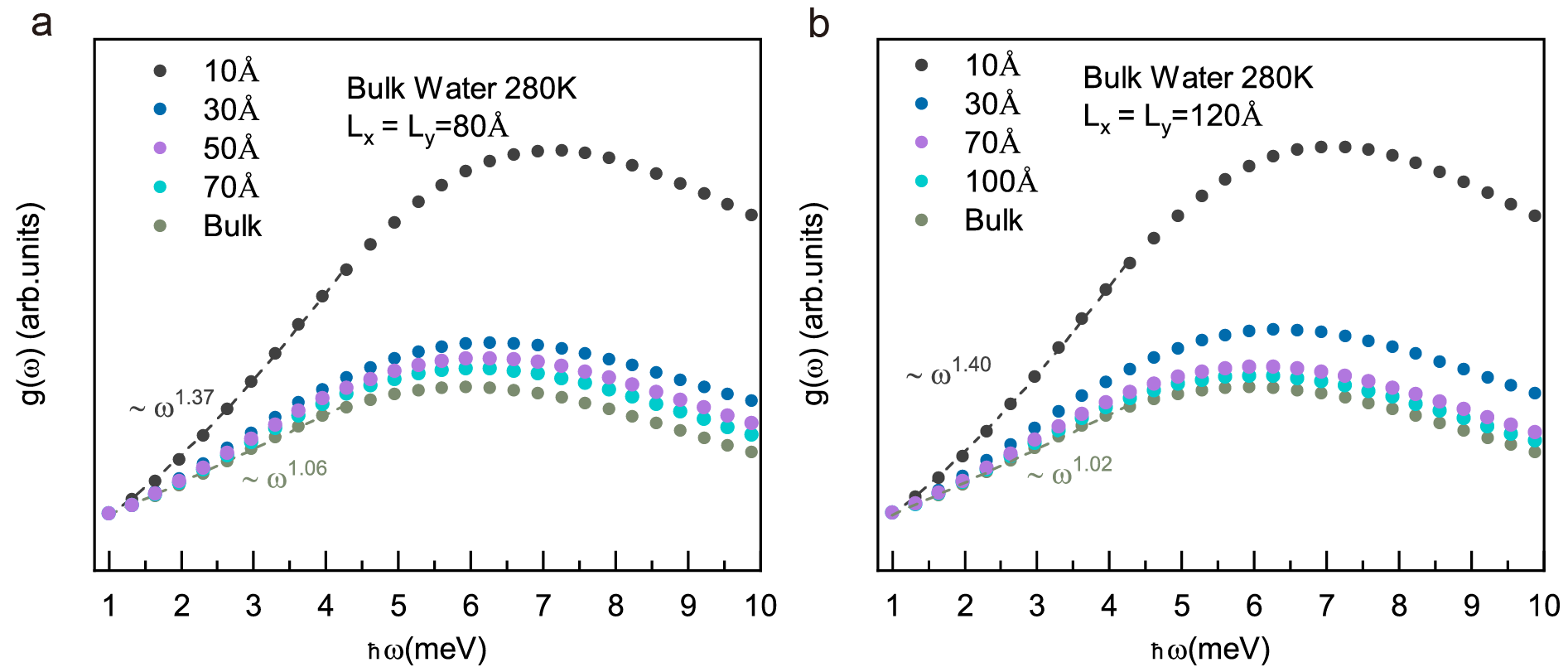}
\caption{The simulated low-frequency DOS for confined water with different confinement sizes when \textbf{(a)}  $L_x=L_y=80 \angstrom$ and \textbf{(b)}  $L_x=L_y=120 \angstrom$.}
\label{fig:size}
\end{figure}

\begin{figure}
    \centering
    \includegraphics[width=\linewidth]{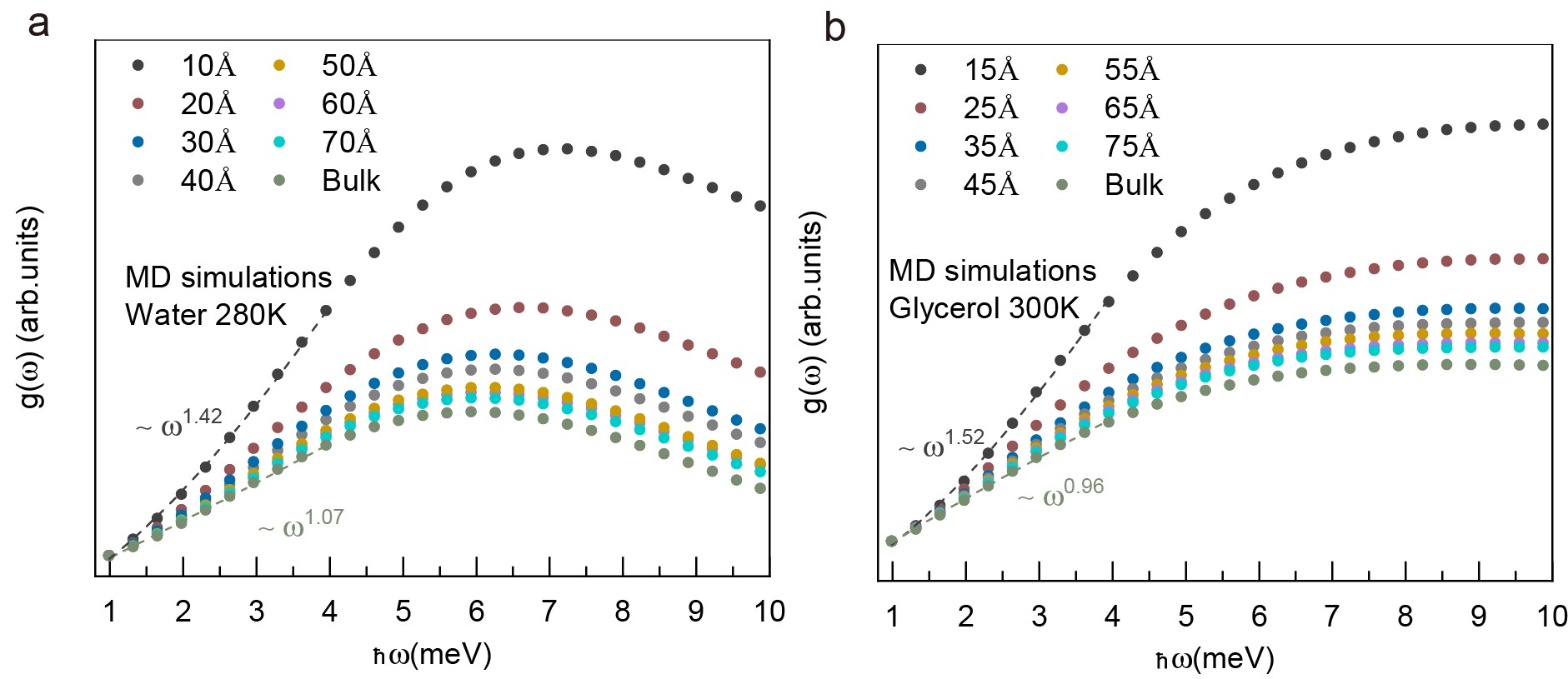}
    \caption{The density of states calculated by all atoms for water \textbf{(a)} and glycerol \textbf{(b)} with different confinement sizes.}
    \label{fig:simsi}
\end{figure}

\begin{figure}
    \centering
    \includegraphics[width=\linewidth]{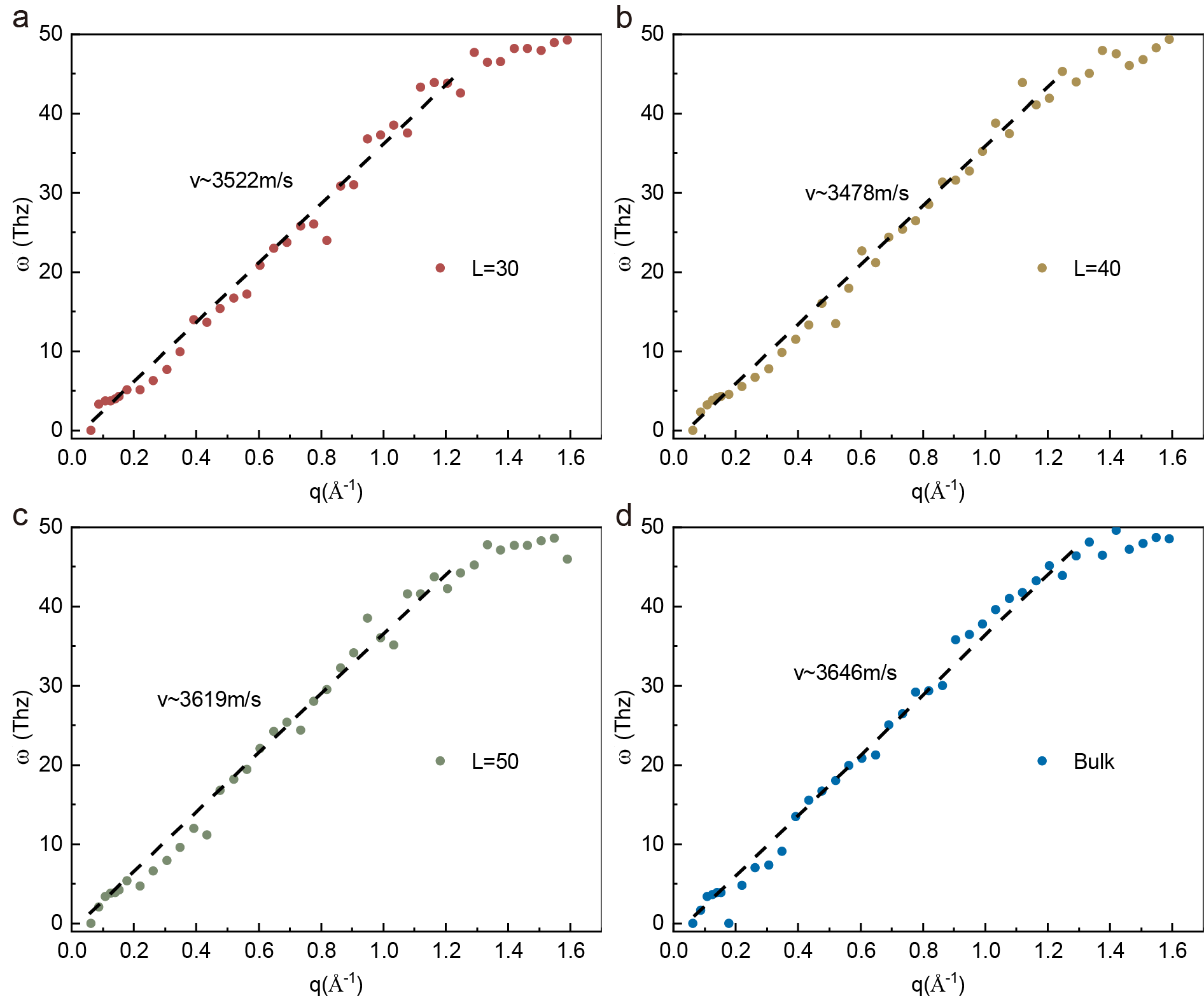}
    \caption{The dispersion relation of longitudinal sound waves in confined water for different confinement sizes: \textbf{(a)}  $30 \angstrom$ , \textbf{(b)}  $40 \angstrom$ , \textbf{(c)} $50 \angstrom$ , and \textbf{(d)} bulk water as a reference.}
    \label{fig:e4}
\end{figure}

\subsection{Longitudinal and shear waves dispersion relations and fits}
In Figure \ref{fig:5} in the main text, we have reported the simulation dispersion relation for collective shear waves in confined water. We have fitted the simulation data with the following formula obtained from k-gap theory:
\begin{equation}
    \mathrm{Re}(\omega)=\sqrt{v^2k^2-\frac{1}{4\tau_g^2}}\equiv v \sqrt{k^2-k_g^2}\label{fifi}
\end{equation}
and extracted the value of the phenomenological parameters $v,\tau_g,k_g$ as a function of the confinement size $L$. The results of the fit are reported in Table \ref{table:2}. The data for the k-gap $k_g$ and the relaxation time $\tau_g$ as a function of $L$ are shown in Figure \ref{fig:4} in the main text. Upon confinement, the propagation speed of the collective shear waves varies only of $7 \%$ and it can therefore be considered in first approximation as constant.\\
Unlike shear waves, the dispersion relation of longitudinal waves in confined water is not significantly affected by the variations in the confinement size. The longitudinal speed of sound varies only of $\approx 3.5 \%$ under confinement and it can be approximated as well by a constant.\\

\subsection{The dynamics of shear waves in liquids and the k-gap model}\label{apptheory}
Let us return to our initial telegrapher equation
\begin{equation}
    \omega^2+i \frac{\omega}{\tau_g}=v^2k^2,
\end{equation}
which describes the dynamical crossover between a propagating wave at large frequency/wave-vector to a overdamped diffusive motion at low energy. By keeping the wave-vector real and assuming the frequency to be a complex number, the solution of the above equation is given by
\begin{equation}
    \omega=-\frac{i}{2\tau_g}\pm \sqrt{v^2k^2-\frac{1}{4\tau_g^2}}\,.
\end{equation}
Clearly, the real part of the frequency becomes non-zero only above the so-called k-gap wave-vector which is given by
\begin{equation}
    k_g=\frac{1}{2v\tau_g}\,.
\end{equation}
Below the k-gap, the mode with lowest energy is diffusive, with dispersion relation $\omega=-i v^2 \tau_g k^2$. Importantly, and often overlooked in the related literature, the appearance of a non-zero real frequency is not enough to have a propagating mode. Indeed, in order to have a well-defined propagating excitation the real part of the frequency must be larger than its imaginary part. This is nothing else than requiring the relaxation time of the wave excitation to be larger than the period of oscillation, such that oscillations might appear before being completely attenuated. In the opposite scenario, where the imaginary part is too large, the wave decays too quickly and no wave-like oscillation has time to take place. By identifying the overdamped-to-underdamped crossover with the location at which imaginary and real part are of the same order, one obtains a critical wave-vector/frequency which are given respectively by
\begin{equation}
    k^*=\frac{1}{\sqrt{2}v\tau_g}\,,\qquad \omega^*=\frac{1}{2\tau_g}.
\end{equation}
Notice that this frequency cutoff is a factor $1/2$ smaller than the original Frenkel criterion $\omega>\omega_F\equiv \tau_g^{-1}$. The latter corresponds to a larger value of the wave-vector, given by
\begin{equation}
    k_F=\frac{\sqrt{5}}{2 v \tau_g}.
\end{equation}

\begin{table}[!h]
\centering
\begin{tabular}{|c|c|c|c|c|c|c|c|c|c|} 
\hline
{$L\,(\angstrom)$}& {30}& {40}& {50}& {60}&{70}& {80}& {90}& {100}\\\hline
$k_g \,(\angstrom^{-1})$ & 0.06283 & 0.08886 & 0.1405 & 0.1405& 0.1539 & 0.1539 & 0.1777 & 0.2205\\\hline
$v$ (m/s)& 1001 & 1011 & 1047 & 1035& 1070 & 1070 & 1071 & 1072\\\hline
$\tau_g \,(ps^{-1})$ & 7.949 & 5.565 & 3.399 & 3.438& 3.036 & 3.036 & 2.626 & 2.114\\
\hline
\end{tabular}
\caption{The results of the fit for the dispersion relation of the collective shear waves shown in Figure \ref{fig:5}a using Eq.\eqref{fifi}.}  
\label{table:2} 
\end{table}

\begin{figure}
    \centering
    \includegraphics[width=0.8\linewidth]{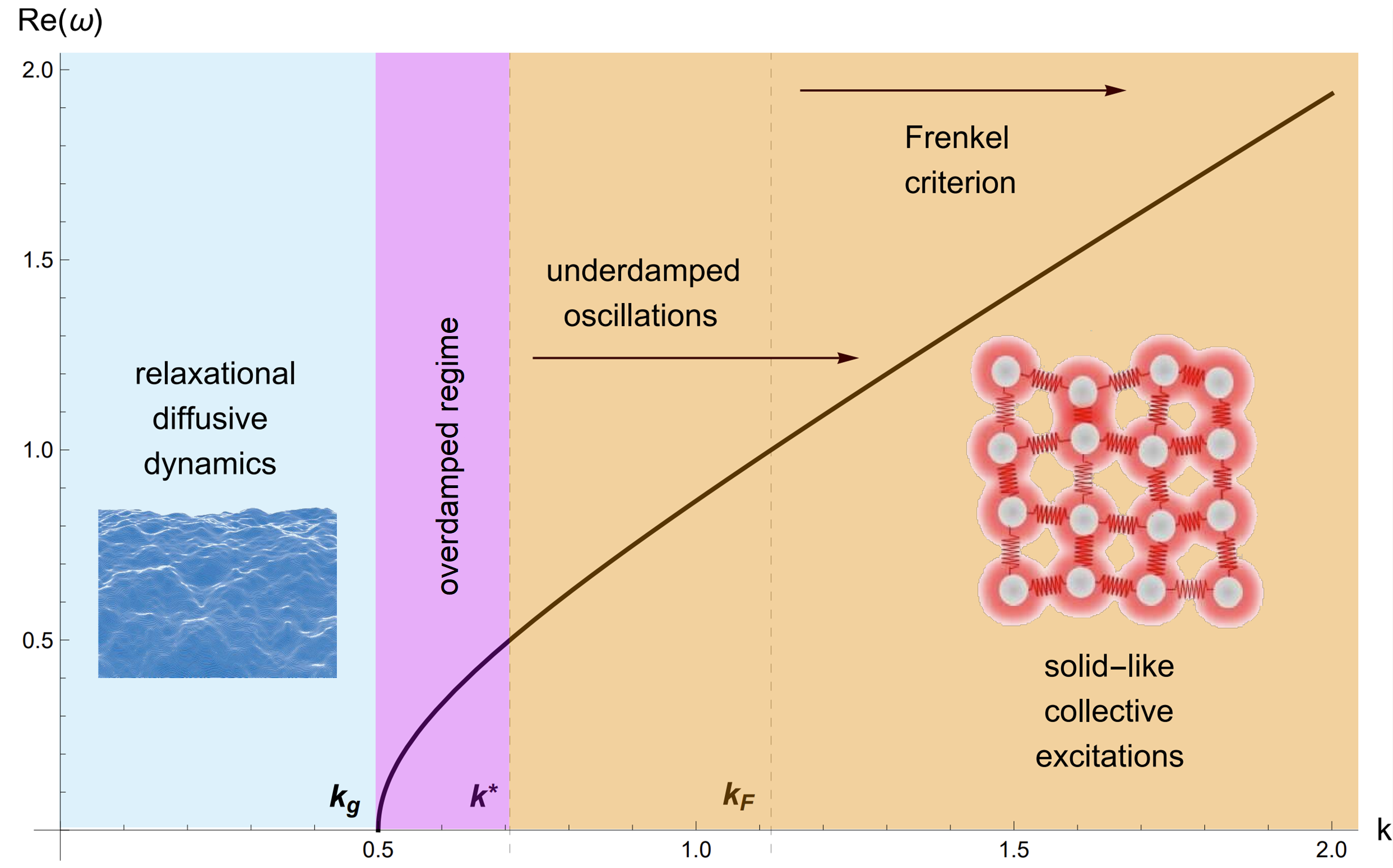}
    \caption{The dispersion relation of the collective waves arising from the telegrapher equation. The different colors indicate the crossover between relaxational diffusive dynamics (with $\mathrm{Re}(\omega)=0$), overdamped oscillations (with $\mathrm{Re}(\omega)<\mathrm{Im}(\omega)$) and underdamped oscillations (with $\mathrm{Re}(\omega)>\mathrm{Im}(\omega)$). For illustration, also the Frenkel criterion $\omega>1/\tau_g$ is illustrated. Here, we have taken $v=\tau_g=1$.}
    \label{fig:app1}
\end{figure}

This hierarchy of scales is displayed in Figure \ref{fig:app1} together with the real part of the dispersion relation of the collective waves. Importantly, apart from $\mathcal{O}(1)$ numerical factor (see for example the factor of $2$ between $\tau_g$ for bulk water and the values reported in the literature for the structural relaxation time \cite{PhysRevLett.82.775,MALOMUZH2019111413,PhysRevFluids.4.123302}), which given the simplicity of the models are likely irrelevant, the major outcome of this analysis is that in classical liquids one can define a critical scale, $L_c \approx 2 \pi v \tau$, below which the dynamics is expected to be solid-like.\\
After discussing the dispersion relation of the collective shear waves described via the telegrapher equation, let us briefly emphasize an interesting outcome of this toy model (see \cite{trachenko2023viscosity} for a recent preprint appeared during the completion of our manuscript and discussing a similar idea). Let us assume that the wave-vector phase space of the shear waves in a liquids is still given by a Debye sphere, as for phonons in solids, and let us assume that in first approximation the effects of confinement on the phase space are negligible. Then, the density of states can be derived using the identity
\begin{equation}
    g(\omega)d\omega= V_k 4\pi k^2 dk,
\end{equation}
where $V_k$ is the volume in $k$-space. By neglecting the imaginary part of the frequency, and identifying $\omega$ with $\mathrm{Re}(\omega)$, we can immediately obtain the frequency dependent density of states which reads
\begin{equation}
    g(\omega)=\frac{2 \pi V_k \omega \sqrt{4 \omega^2+\gamma^2}}{v^3}\qquad \text{where}\qquad \gamma \equiv 1/\tau_g\,.
\end{equation}
\begin{figure}
    \centering
    \includegraphics[width=0.8\linewidth]{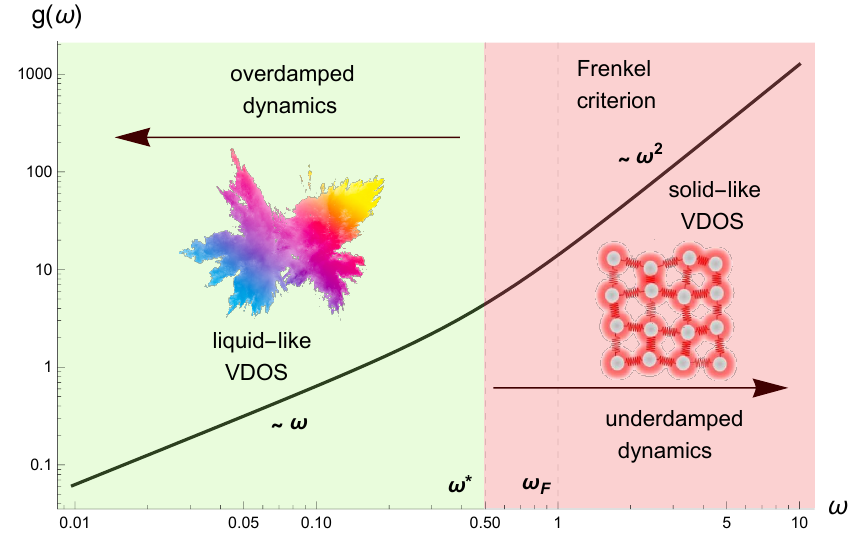}
    \caption{The density of states which can be naively derived from a wave-like excitation whose dispersion follows from the telegrapher equation. The DOS displays a crossover between a liquid-like linear behavior at low energy to a Debye law at large frequency. The crossover corresponds to the scale at which the collective excitation passes from being overdamped to being underdamped which approximately corresponds to the Frenkel frequency $\omega_F\equiv 1/\tau_g$.}
    \label{fig:app2}
\end{figure}

The behavior of this DOS is shown in Figure \ref{fig:app2}. The DOS displays a crossover between a low frequency linear scaling $g(\omega) \propto \omega$ to a quadratic Debye law at large frequency. Interestingly, the change in the behavior coincides exactly with the crossover between the overdamped to underdamped dynamics. Not surprisingly, when the wave is underdamped, the corresponding DOS is Debye-like as for solids. On the contrary, whenever the wave is overdamped or even not-well defined (relaxational), the DOS is liquid like. At least in first approximation, this simple computation tells us that the liquid-like linear behavior in liquids should disappear, and smoothly connect with the quadratic Debye scaling, at a frequency scale which is approximately of the order of the Frenkel frequency $\omega_F\equiv 1/\tau_g$, if $\tau=\tau_g$. In other words, for $\omega \gg 1/\tau_g$, even a liquid is expected to display a Debye-like density of state $g(\omega) \propto \omega^2$. When the temperature is high, then $1/\tau$ is very large and the Debye solid like scaling is likely overwhelmed and dominated by other frequency modes. Nevertheless, when the temperature is low, or like in our case when confinement makes $1/\tau_g$ small enough, then one could possibly directly see the crossover between $\omega$ and $\omega^2$ in the low-frequency DOS of liquids. This is exactly what we show in the main text in Figure \ref{fig:5} for the experimental data of confined water. This is qualitatively in perfect agreement with k-gap theory and the original Frenkel idea.
\section*{Data Availability Statement}
The datasets generated and analysed during the current study are available upon reasonable request by contacting the corresponding authors.

\providecommand{\latin}[1]{#1}
\makeatletter
\providecommand{\doi}
  {\begingroup\let\do\@makeother\dospecials
  \catcode`\{=1 \catcode`\}=2 \doi@aux}
\providecommand{\doi@aux}[1]{\endgroup\texttt{#1}}
\makeatother
\providecommand*\mcitethebibliography{\thebibliography}
\csname @ifundefined\endcsname{endmcitethebibliography}  {\let\endmcitethebibliography\endthebibliography}{}

\end{document}